\documentclass[review, authoryear]{elsarticle}
\usepackage[utf8]{inputenc}
\usepackage{todonotes}
\usepackage{float}
\usepackage[nolist]{acronym}
\usepackage{comment}
\usepackage{amsmath,amssymb,amsfonts}

\usepackage{array}
\newcolumntype{H}{>{\setbox0=\hbox\bgroup}c<{\egroup}@{}}
\newcolumntype{P}[1]{>{\centering\arraybackslash}p{#1}}
\newcolumntype{L}[1]{>{\RaggedRight\hspace{0pt}}p{#1}}
\newcolumntype{R}[1]{>{\RaggedLeft\hspace{0pt}}p{#1}}
\usepackage{longtable}
\usepackage{multirow}
\usepackage{pbox}
\usepackage{rotating}
\usepackage{makecell}
\usepackage{graphicx}
\usepackage{booktabs}
\usepackage[bookmarks,hidelinks]{hyperref}
\usepackage{xurl}
\usepackage{pdfpages}

\usepackage[width=15cm]{geometry}
\usepackage{floatflt,epsfig, wrapfig}

\usepackage{natbib}

\setcitestyle{authoryear,open={[},close={]}}

\usepackage{lineno}
\modulolinenumbers[5]

\journal{Information \& Management}

\definecolor{ocolor}{rgb}{1,0,0.4}

\renewcommand{\arraystretch}{1.25}

\usepackage{pdfrender}

\usepackage{pifont}

\journal{Information \& Management}

\title{Designing a Framework for
Digital KYC Processes Built on Blockchain-Based Self-Sovereign Identity}

\date{\today}

\author{\texorpdfstring{\mbox{Vincent Schlatt\,{$^{\mathrm{a},\mathrm{b},\dagger,\ast}$}}}}
\author{\texorpdfstring{\mbox{Johannes Sedlmeir\,{$^{\mathrm{a},\mathrm{b},\dagger}$}}}}
\author{\texorpdfstring{\mbox{Simon Feulner\,{$^{\mathrm{b},\mathrm{c}}$}}}}
\author{\texorpdfstring{\mbox{Nils Urbach\,{$^{\mathrm{b},\mathrm{c}}$}}}}

\address{
    $^\mathrm{a}$
    FIM Research Center, University of Bayreuth, Bayreuth, Germany\\
    $^\mathrm{b}$
    Project Group Business \& Information Systems Engineering of the Fraunhofer FIT, Bayreuth, Germany\\
    $^\mathrm{c}$
    Frankfurt University of Applied Sciences, Frankfurt, Germany\\~\\
    $^\dagger$ 
    Both first authors contributed equally to the article \\
    $^\ast$
    Corresponding author: vincent.schlatt@fim-rc.de
    \\~\\~\\~\\~\\~\\
    \large{\textup{This is the accepted version of \url{https://doi.org/10.1016/j.im.2021.103553}, published in the Special Issue ``Blockchain Innovations: Business Opportunities and Management Challenges'' in Information \& Management.}}
}

\begin{document}
\begin{frontmatter}

\begin{abstract}
Know your customer (KYC) processes place a great burden on banks, because they are costly, inefficient, and inconvenient for customers. While blockchain technology is often mentioned as a potential solution, it is not clear how to use the technology's advantages without violating data protection regulations and customer privacy.
We demonstrate how blockchain-based self-sovereign identity (SSI) can solve the challenges of KYC. We follow a rigorous  design science research approach to create a framework that utilizes SSI in the KYC process, deriving nascent design principles that theorize on blockchain's role for SSI.
\\
\end{abstract}


\end{frontmatter}

\clearpage
\pagenumbering{gobble}
\textbf{Highlights}
\begin{itemize}
    \item SSI can make KYC processes completely digital, efficient, compliant, and convenient.
    \item No personal data need to be stored on a blockchain.
    \item SSI inhibits data silos, lock-in effects, and aggregation of market power.
    \item Blockchain's role for SSI should be more restricted than is often proposed.
\end{itemize}

\begin{figure}[H]
    \centering
    \includegraphics[width=\linewidth, trim=0.5cm 0.5cm 0.5cm 0.5cm, clip]{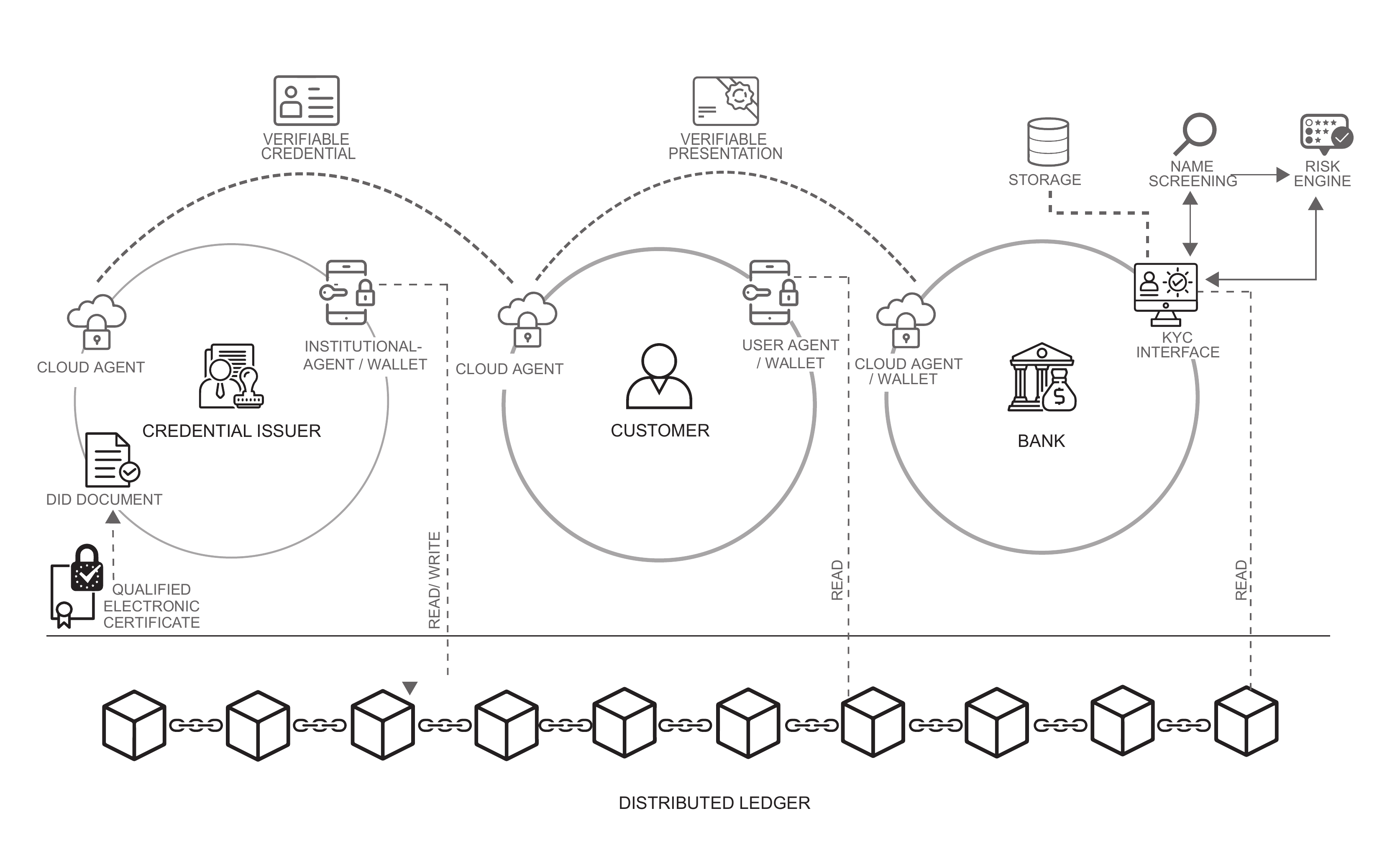}
    \label{fig:abstract}
\end{figure} 
~\\

\begin{acronym}
\setlength{\itemsep}{-0.3cm}
\acro{ADR}[ADR]{action design research}
\acro{AFI}[AFI]{Alliance for Financial Inclusion}
\acro{AML}[AML]{anti-money laundering}
\acro{BIE}[BIE]{building, intervention, and evaluation}
\acro{CA}[CA]{certificate authority}
\acro{CFT}[CFT]{counter-terrorism financing}
\acro{DKMS}[DKMS]{decentralized key management system}
\acro{DNS}[DNS]{domain name system}
\acro{DID}[DID]{decentralized identifier}
\acro{DIF}[DIF]{Decentralized Identity Foundation}
\acro{DLT}[DLT]{distributed ledger technology}
\acro{DP}[DP]{design principle}
\acro{DPKI}[DPKI]{decentralized public key infrastructure}
\acro{DSR}[DSR]{design science research}
\acro{EGIZ}[EGIZ]{E-Government Innovationszentrum}
\acro{eKYC}[eKYC]{electronic KYC}
\acro{eIDAS}[eIDAS]{electronic Identification, Authentication, and Trust Services}
\acro{EU}{European Union}
\acro{eSSIF}[eSSIF]{European Self-Sovereign Identity Framework}
\acro{FATF}[FATF]{Financial Action Task Force on Money Laundering}
\acro{FEDS}[FEDS]{framework for evaluation in design science}
\acro{GDPR}[GDPR]{General Data Protection Regulation}
\acro{GLEIF}[GLEIF]{Global Legal Entity Identifier Foundation}
\acro{IMA}[IMA]{Money Laundering Act}
\acro{IMS}[IMS]{identity management system}
\acro{IoT}[IoT]{Internet of Things}
\acro{KYC}[KYC]{know your customer}
\acro{MLA}{Money Laundering Act}
\acro{P2P}[P2P]{peer-to-peer}
\acro{PKI}[PKI]{public key infrastructure}
\acro{RBA}[RBA]{risk-based approach}
\acro{SSI}[SSI]{self-sovereign identity}
\acro{SSO}[SSO]{single sign-on}
\acro{ToIP}[ToIP]{Trust over IP}
\acro{UIDAI}[UIDAI]{Unique Identification Authority of India}
\acro{VC}[VC]{verifiable credential}
\acro{VP}[VP]{verifiable presentation}
\acro{W3C}[W3C]{World Wide Web Consortium}
\acro{WEF}[WEF]{World Economic Forum}
\acro{WoT}[WoT]{Web of Trust}
\acro{ZKP}[ZKP]{zero-knowledge proof}
\end{acronym}
\clearpage 

\pagenumbering{arabic}
\setcounter{page}{1}

\section{Introduction}
\acresetall
\label{sec:introduction}
Financial regulation has three primary goals: financial inclusion, financial stability, and market integrity~\citep{zetzsche2018digital}. To achieve the goal of market integrity, regulators have introduced several regulatory requirements into the financial sector, such as the \ac{FATF} recommendations, which seek to prevent money laundering and the financing of international terrorism, as well as Basel III, in reaction to the global financial crisis in 2008~\citep{arner2016emergence}.
To remain compliant with this regulatory regime, financial institutions must perform in-depth due diligence to identify their customers and to understand the purpose of their activities, a process formally known as \ac{KYC}~\citep{arasa2015determinants}, in which
customers typically need to be physically present at the bank's branch or on a video call to provide personally identifying information, such as a passport or an ID card.

This process is problematic for banks, because it is cost-intensive, time-consuming, and inconvenient for customers~\citep{zetzsche2018digital}.
Thus, there have been several attempts at improvement, mostly involving the digitization of particular process steps. For instance, some banks use their customers' analog proof of identity, such as passports, and create internally used digital customer identities to improve the process flow. However, this approach again suffers from inefficiencies, since it is error-prone, time-consuming~\citep{jessel2018digital}, and highly repetitive~\citep{zetzsche2018digital}. The lack of shared standards and banks' reservations about sharing customer information with competitors also limit the reusability of a customer's \ac{KYC} data at different banks~\citep{arner2019identity}.

A central utility that collects and provides identity-related data for an \ac{eKYC} process, as in India or Australia, is often mentioned as a solution to the aforementioned problems~\citep{zetzsche2018digital, arner2019identity, perlman2019focus}, since it can reduce costs and significantly shorten \ac{KYC} onboarding processes~\citep{rajput2017towards}. However, recent reports of leaks and misuses of personal data have lowered the confidence of both banks and customers in solutions that involve creating central data silos~\citep{swinhoe2020breaches}. Moreover, there are jurisdictions in which such a centralized service run by the government is not feasible~\citep{rieger2019building}. Generally, the fear that such a distinct service provider will aggregate significant market or political power impedes the establishment of a widely accepted centralized service provider~\citep{zavolokina2020management}.

Thus, both researchers and practitioners have identified blockchain technology as a potential solution to the latter problems. Blockchains can provide neutral platforms for digital cross-organizational workflows~\citep{guggenberger2020improving}, mitigating the threat of market power aggregation. At the same time, blockchain technology enables digital trust through synchronized redundancy and therefore transparency, tamper-resistance, and enforcement of processes through smart contracts~\citep{rossi2019blockchain}. However, it is well known that blockchain technology's built-in transparency and append-only structure aggravates privacy-related problems~\citep{rieger2019building}. Particularly, the European \ac{GDPR} grants individuals the \emph{right to be forgotten}, which means that they can demand that their private data be deleted at any time as soon as the purpose for their storage has expired. As data stored on a blockchain practically cannot be erased, implementations such as~\citeauthor{moyano2017kyc}'s~[\citeyear{moyano2017kyc}] where \ac{eKYC}-related information is stored transparently on-chain, are not a viable solution.

As an alternative, one could think of depositing the \ac{KYC} information in a standardized way at the one and only entity involved in each of its \ac{KYC} processes -- the customer. These considerations lead to the concept of \ac{SSI}, which seeks to establish holistic digital identity management on the paradigm that a user controls all their data and attestations, similar to today's analog identity management via a system of plastic cards in physical wallets. Yet, \ac{SSI} is still strongly linked with blockchain technology because it requires a neutral platform that provides governance, standards, and essential public information to check the validity of attestations. This goal of an interoperable digital identity management system without a distinct central authority in control makes \ac{SSI} very attractive for digitizing the \ac{KYC} process.

A recent pilot in the UK that investigated the opportunities of \ac{SSI}-based \ac{KYC} found that an \ac{SSI}-based ``portable identity significantly improves both consumer experience and protection, while accelerating customer onboarding and reducing KYC and compliance-related costs for financial institutions''~\citep{ledgerinsights2020ssikyc}.
While research on the problem and approaches to \ac{SSI}-based \ac{eKYC} onboarding have recently emerged~\citep{soltani2018new}, they have not covered topics such as user orientation, coverage of the entire \ac{KYC} process, or platform independence. Further, \citet{soltani2018new} focused on implementing the principles of \ac{SSI} without acknowledging that \ac{SSI} is a tool to achieve an improved \ac{KYC} process from the perspective of banks. Looking at \ac{SSI} generally, in the related literature, blockchain's role in this context remains largely unclear.
Thus, both research and practice need a generic and validated framework that guides the design of \ac{SSI} solutions for entire \ac{eKYC} processes and an overview of the resulting implications to assess the potential benefits and to learn how to leverage them. Further, we still lack generic \acp{DP} to guide the development of \ac{SSI} solutions based on blockchain technology that can also be used in other sectors \citep{liu2020design}. We seek to design a framework for an \ac{eKYC} process built on blockchain-based \ac{SSI} and to derive initial generic \acp{DP}.
We develop and evaluate our framework in a rigorous \ac{DSR} approach, incorporating both existing theoretical knowledge and practitioners' perspectives through semi-structured expert interviews. Thus, we extend the literature on \ac{eKYC} by providing a comprehensive architecture and process framework, discussing the roles of blockchain and \ac{SSI} for \ac{eKYC}, and producing generalizable knowledge on the design, opportunities, and challenges of blockchain-based \ac{SSI} systems. The \acp{DP} we develop from our \ac{DSR} suggest that blockchain's role in \ac{SSI} should be more restrictive than is typically proposed in the literature in order to make systems scalable and compliant with regulations. We also guide practitioners on how to design the respective systems.

The remainder of this study is structured as follows: In Section~\ref{sec:background}, we present background knowledge on \ac{KYC} processes, blockchain technology, and \ac{SSI} that is necessary to understand the work that follows. In Section~\ref{sec:methods}, we present our \ac{DSR} method. In Section~\ref{sec:objectives}, we derive objectives for the \ac{eKYC} framework and evaluate them through expert interviews. We present the framework, including the \ac{SSI}-based \ac{eKYC} architecture and process, in Section~\ref{sec:framework}. Section~\ref{sec:evaluation} continues with the evaluation of the framework along the derived objectives. In Section~\ref{sec:discussion}, we discuss the findings, develop nascent \acp{DP} for blockchain-based \ac{SSI}, and provide managerial and theoretical implications. In Section~\ref{sec:conclusion}, we summarize our results, identify limitations, and provide an outline for further research.

\section{Background}
\label{sec:background}

\subsection{The KYC process and centralized attempts at eKYC}
\label{subsec:KYC}
After the original \ac{FATF} Forty Recommendations were drawn up in 1990, they were revised in 1996 to account for the latest money laundering techniques. These recommendations for \ac{AML} have been adopted by more than 130 countries and are therefore considered to be the international standards~\citep{ruce2011anti}. A key element of these recommendations is the \ac{KYC} process. Financial institutions are urged not to open anonymous accounts or accounts with obviously fictitious names. In this context, due diligence is recommended to verify the identity of customers through independent, credible documents. The purpose of the business relationship must also be verified. Further, the \ac{KYC} process should include ongoing monitoring of transactions to identify suspicious customer behavior~\citep{force2004fatf}.

\begin{figure}[!b]
    \centering
    \includegraphics[width=0.8\linewidth]{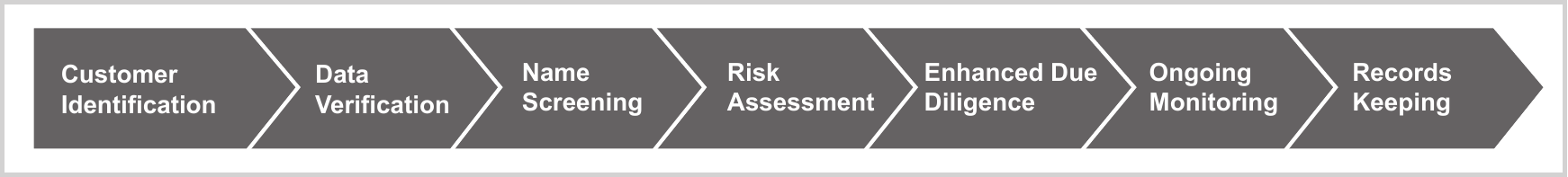}
    \caption{The \ac{KYC} process.}
    \label{fig:kyc}
\end{figure}

KYC processes may differ, owing to countries' different regulatory requirements and the banks' specific requirements. However, some repeating core activities of the \ac{KYC} process can be identified (see Figure~\ref{fig:kyc}). The process begins with the collection of data about potential customers to identify them. Government-issued documents such as ID cards, driver's licenses, or passports are preferred. Documents from other companies in the financial sector, as well as other documents relevant for the identification of persons, such as telephone or gas invoices, can also be used~\citep{mugarura2014customer}.
After a customer is identified and the identity data claims are verified, the bank checks whether the person represents a risk for the financial institution. This includes matching against a list of known terrorists, criminals, and politically exposed persons~\citep{arasa2015determinants}. Further initial and ongoing measures follow to allow the bank to do permanent risk monitoring.

The process of initial verification and ongoing monitoring of activities must be repeated for each customer, and every customer must undergo this process again when opening an account with a new bank. Thus, the \ac{KYC} process is very time-intensive and inconvenient for both customers and banks, resulting in poor customer experiences and fewer account openings. For instance, 89\,\% of surveyed customers did not have a good experience with the \ac{KYC} process~\citep{reuters2016kyc}, and criticized the onboarding process because it was time-intensive and involved posting several documents. To avoid losing their customers and revenue opportunities, financial institutions must make essential improvements. Also, the overall market efficiency could benefit from enhanced competition owing to lower switching costs.
Further, the high effort required for this process and the lack of automation of some manual steps result in high costs for the financial institutions. A survey of 800 financial institutions found that the annual cost for \ac{KYC} per bank is approximately USD \$60~million~\citep{reuters2016kyc}.

The primary focus and motivator of regulatory efforts toward \ac{KYC} is the avoidance of money laundering through financial institutions. Failure to comply with regulatory requirements may further increase \ac{KYC} process cost through considerable fines~\citep{moyano2017kyc}. A major goal in \ac{KYC} efforts for financial institutions is often, therefore, the avoidance of fines or loss of reputation, at preferably low ownership costs. However, some institutions also see \ac{KYC} as an opportunity, since it enables them to better understand customers, identify their needs and behaviors, create customized products, and improve customer relationships, ultimately leading to higher company profits~\citep{ruce2011anti}.

The key to simultaneously reducing compliance costs, preventing regulatory penalties, and harnessing new potential lies in the digitization and automation of processes and the resulting opportunities for data processing and analysis~\citep{lootsma2017blockchain}.
One often used approach to improve the \ac{KYC} process is the digitization of analog ID documents, which typically involves some facial verification step by a combination of human and machine learning examination. One step further are approaches that seek to abandon analog documents altogether, which is why the term \ac{eKYC} is often used here~\citep{christie2018setting}. A sector-wide \ac{eKYC} utility could avoid the repeated execution of the \ac{KYC} process at different banks. These systems typically use biometrics such as fingerprints, iris scans, or facial recognition. The data are then stored on a smart ID card and online in a central database, together with personally identifiable information such as the customer's name, age, and place of residence. During the \ac{KYC} process, the customer's biometric data are captured and matched against the data in the central online database.

An example of such a sector-wide eKYC utility is India's Aadhaar system. Indian citizens must provide various demographic and biometric data~\citep{zetzsche2018digital}, which are stored in a central database. The system has led to much faster onboarding times and to fewer losses from fraud and corruption~\citep{economist2016aadhaar}. However, several data breaches have raised questions regarding the privacy and security of the system~\citep{zetzsche2018digital}.
Further, if operated by a public authority or heavily regulated, identity systems such as India's could be used by governments for mass surveillance without citizens' knowledge. On the other hand, on an international level or if operated by a private company, threats of monopolies and market power may keep banks from participating in such a system~\citep{zavolokina2020management}. Thus, while it is a key step toward increasing process efficiency, the design of identity systems is critical for their success and acceptance, because this involves the management of highly sensitive data and misuse should be prevented. Centralized databases to store users' personal data are attractive targets for attackers who can steal large amounts of sensitive information. Consequently, they are challenging to secure~\citep{sedlmeir2021identities}. Similar concerns regarding the creation of centralized service providers for non-competitive data arise in various further contexts beyond \ac{KYC}. For the \ac{KYC} procedure, digital identity systems must therefore be considered from the perspectives not only of efficiency and user experience but also privacy and security~\citep{perlman2019focus}.

\subsection{Blockchain technology and decentralized approaches to eKYC}
\label{subsec:DLT}
Owing to the limitations and downsides of centralized platforms, banks have started looking for alternatives, one being \ac{DLT}. Key components of \ac{DLT} are a peer-to-peer network where all data are replicated across multiple peers, and an associated consensus protocol operated by specific nodes to ensure the validity of state modifications (\emph{transactions}) and to synchronize all replicas~\citep{glaser2017pervasive, kolb2020core}. Authentication on \ac{DLT} is conducted through public key cryptography, which allows one to participate in consensus or to interact with the network and authorize transactions. Distributed ledgers are resistant to crashes and even the malicious behavior of a small subset of nodes, making them a highly available and decentralized digital infrastructure. However, \ac{DLT} also has considerable drawbacks concerning scalability and privacy, owing to the redundant operation of all transactions~\citep{kolb2020core}.
Blockchains\footnote{We use the terms blockchain and \ac{DLT} interchangeably in this work} are a special case of \ac{DLT}, and are probably the most widely used. The key characteristic of blockchain architectures is that transactions are batched into blocks, and each block of data contains the previous block's hash value. The blocks therefore form an append-only structure (\emph{chain}) with the aim of establishing a tamper-resistant historical record~\citep{butijn2020blockchains}.

Thus, blockchains can serve as a physically decentralized yet logically centralized source of truth for information, making them suitable for decentralized asset management~\citep{rossi2019blockchain}. Guaranteeing transparency and the enforcement of rules while ensuring the independence from a distinct node can be major advantages of blockchain solutions for cross-organizational workflow management \citep{fridgen2018cross}.
Businesses and public authorities have realized \ac{DLT}'s potential for the digitization of their cross-organizational processes, leading to a large number of projects~\citep{casino2019systematic}.
Considering the aforementioned generic properties of \ac{DLT}, a blockchain-based neutral platform on which banks could collaborate on \ac{eKYC} seemed very appealing, since this approach can eliminate the threat of monopolies. However, it aggravates privacy-related problems, since tamper resistance and redundancy imply not only that stored on-chain data are visible to all nodes but also that it is practically impossible to delete on-chain data~\citep{rieger2019building,kolb2020core}. It therefore does not make sense to store personal data on the ledger~\citep{dunphy2018first} and doing so contradicts regulation such as the \ac{GDPR}, which includes the \emph{right to be forgotten}.

This fact significantly complicates the conceptual integration of a \ac{DLT} into the \ac{KYC} process. \citet{moyano2017kyc}, \citet{biryukov2018privacy}, and \citet{norvill2019blockchain}, for instance, proposed writing a proof about the successful completion of the \ac{KYC} process in the form of a hash value on a blockchain. In this concept, the de facto data are still stored in a centralized database operated by banks or a service provider. Once a bank customer has completed the \ac{KYC} process, it will be sufficient for the customer to prove their identity using the hash value in the ledger. Although the efficiency of the process can thus be increased, central parties with full control of and access to the data are still necessary with these approaches, again causing the described security and privacy challenges. Further, challenges regarding the binding of cryptographic keys to customers as well as the management of permissions for exchanging customer data remain, while the benefit of using a blockchain is not yet clear, as a public key infrastructure and certificates based on digital signatures can provide tamper-proof evidence of a completed \ac{KYC} process. \cite{Ostern2020blockchain} acknowledged the challenges of storing customer data on a blockchain in their development of a blockchain-based system for \ac{KYC} to satisfy the requirements of initial coin offerings. Thereby, only the statuses of completed \ac{KYC} processes are stored on a blockchain. However, in their design, the customers' identity data remain with a centralized provider specialized in  \ac{KYC}, and the protocol for exchanging data between the banks and the eKYC provider remains unspecified.

\subsection{SSI and its proposed application to eKYC}
\label{subsec:SSI}
Today, identification and authentication are usually carried out against a service provider using a username and password. The reason for the widespread use of this so-called \emph{centralized} identity model lies in its simple implementation and in the full control of the service providers, who can minimize risks if no third party is involved for authentication. Users also benefit from the fact that they only have to pass on the information necessary for the context in question~\citep{clauss2001identity}. However, the increasing use of internet services has made this system inconvenient for users, since they have to remember the login data for each additional service, and manual input or repeated verification processes of attributes are necessary~\citep{maliki2007identity}. This leads to poor user experiences and security issues, as users tend to reuse passwords across many services. Moreover, service providers need to rely on the validity of the data provided by the customer, which can result in bad data quality and costs for fraud that cannot be traced back to a natural person. Service providers also usually store the data in large data silos -- a popular target for hackers~\citep{rajput2017towards}.

In an attempt to improve user experience, the so-called \emph{federated} identity model was developed~\citep{maler2008venn}. This concept allows for the use of digital identities for authentication and proof of attributes across organizational and system boundaries. An identity provider, such as Facebook or Google, manages users' digital identities and makes them available to relying parties. The fundamental prerequisite for this identity model is the establishment of a trust relationship between the identity provider and the relying party.
Federated identity management improves user experience, since the users no longer have to remember a large number of user names and passwords, and only need a single sign-on~\citep{lim2018blockchain}. However, from the perspective of privacy and security, such services are even more problematic than centralized systems~\citep{maler2008venn}.

If privacy and security need to be improved, there must no longer be any central parties that have access to users' full digital identities and the associated data. Rather, control must be decentralized. By using public key cryptography, users can create their own identifiers -- also known as \acp{DID} -- and prove control over them.
Users can then append information to these identifiers. For contexts in which some attested attributes require confirmation, users can collect credentials from trusted authorities, such as government agencies, companies, or universities~\citep{w3c2019vc}. \acp{DID} and the associated cryptographic keys, as well as credentials, are stored by users in so-called digital wallets, for instance on smartphones, computers, or in the cloud with a provider of their choice. Such a system is comparable to the physical credentials, e.g., plastic cards, we carry in our physical wallets~\citep{DBLP:journals/csm/AvellanedaBBBDD19}. Since users fully control their data, this approach has been called \emph{self-sovereign}~\citep{allen2016path}.

Such an approach requires open-source and open-standard technology~\citep{wagner2018self}.
Various implementations of \ac{SSI} are possible and have been realized, but currently many commonly used implementations build on the \ac{DID} standard being developed by the \ac{W3C}~\citep{w3c2020did}. A \ac{DID} is always associated with a \ac{DID} document that contains information such as public key material used to delegate and prove ownership and control of a \ac{DID}~\citep{w3c2020did}, and to establish a secure (encrypted) communication channel with this \ac{DID}. Besides the purpose of standardization, \acp{DID} create a reference point for bilateral interactions that is portable across domains and does not require a centralized authority to register, resolve, update, or revoke the identifiers~\citep{soltani2018new}.
In this sense, \acp{DID} are not strictly necessary for \ac{SSI}, but provide functionalities that go beyond the mere capabilities of \ac{DPKI}.

\begin{figure}[!htb]
    \centering
    \includegraphics[width=0.65\linewidth]{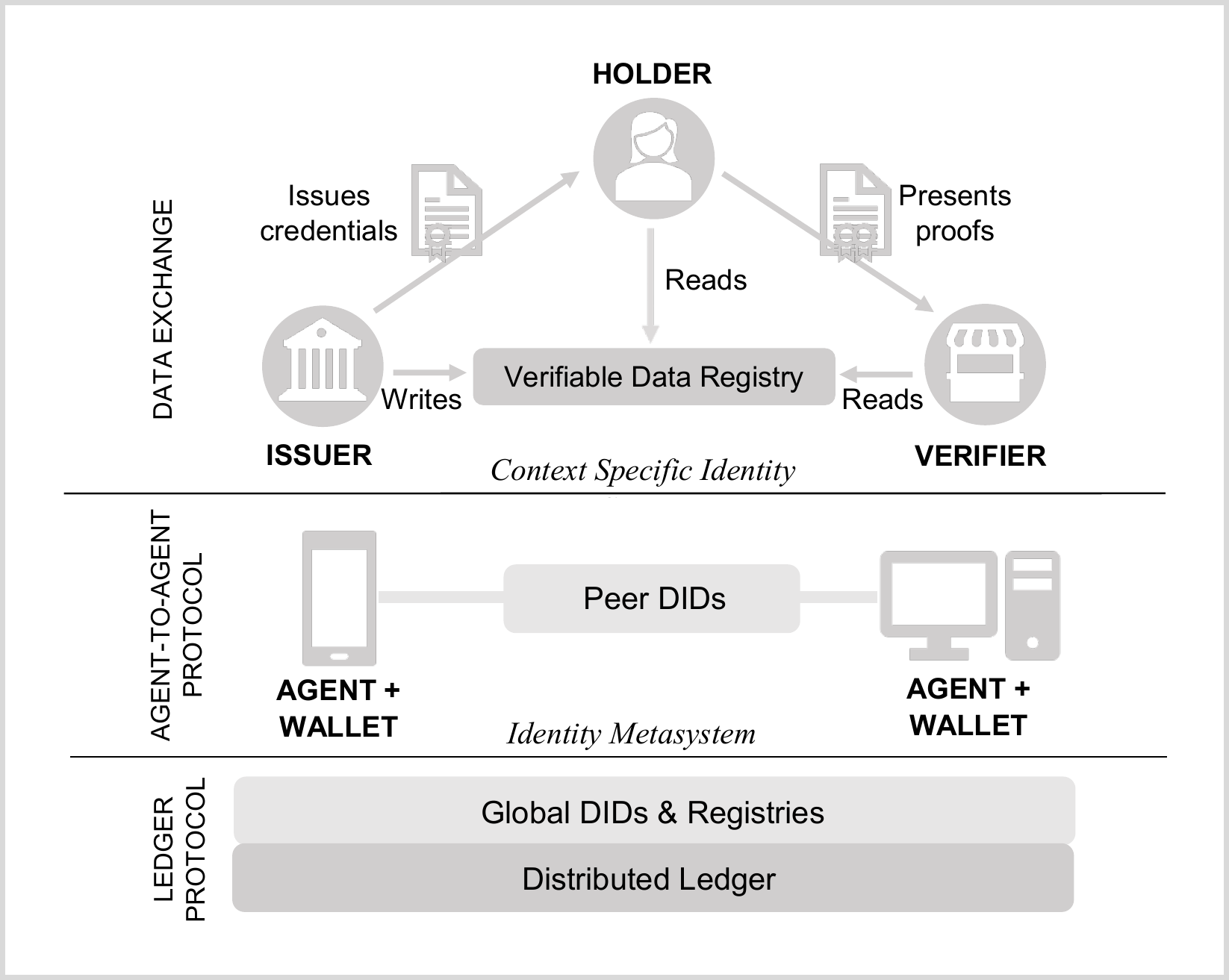}
    \caption{Layers of \ac{SSI}-based identity management based on \citep{ToIP2020Whitepaper}.}
    \label{fig:SSI_Layers}
\end{figure}

Credentials that provide cryptographic evidence of who created them and who they were created for are widely known as digital certificates. A new flavor, called \acp{VC}, is currently the subject of standardization efforts by the \ac{W3C}~\citep{w3c2019vc}. Their validity and whether they have expired or been revoked can be verified without having to communicate with the issuer of the credentials, by checking the issuer's digital signature and a public yet privacy-preserving revocation registry. However, this approach requires an established trust relationship between a verifier and the credential issuer~\citep{muhle2018survey}. The decentralized approach regarding the reliable and trustworthy provision of public information that is necessary to verify \ac{VC} data is enabled by the use of \ac{DLT}. \ac{DLT} acts as a \emph{single point of truth} and thus as a generally acceptable and immutable location for the storage and management of information about standards, issuers of \acp{VC} (e.g., their public signing keys), and revocation status. \ac{DLT} therefore provides a censorship-resistant storage facility for information that must be publicly available, without the need for a central entity such as a certificate authority~\citep{muhle2018survey}. \ac{SSI}'s key roles and building blocks are summarized in Figure~\ref{fig:SSI_Layers}.

Besides the security aspect, a widely acknowledged opportunity of \ac{SSI} is enhanced privacy features~\citep{sedlmeir2021identities}. One the one hand, by default, different identifiers, so-called pairwise \acp{DID} (pseudonyms), can be used in different interactions. Global \acp{DID} are required only for public entities that want to aggregate reputation or trust, such as credential issuers. Further, some implementations of \acp{VC} can prove the correctness of claims, such as the existence of the issuers' signature on the \ac{VC}, without the need to reveal the value of the signature itself or all attributes that are attested on the credential. This significantly mitigates the correlatability of conventional digital certificates by means of their digital signature, and ultimately allows for enhanced privacy while still exchanging the information that is required to build the trust relationship that is necessary for interactions and business~\citep{davie2019trustoverip,hardman2020zkpsavvy}.

Since the concept places users in the center and leaves them in full control, some general challenges arise with \ac{SSI}. First, appropriate measures must be taken to ensure user friendliness. Users must take care of storing the credentials and managing the keys themselves. So-called digital agents or wallets are used for this, either directly on an edge agent (e.g., a smartphone or laptop) or with cloud agents that can only be accessed by the user~\citep{lyon2019identity}. Cloud agents are helpful, since edge agents cannot guarantee permanent online availability~\citep{reed2018dkms}. Further, problems such as recovery in the case of device loss or theft must be addressed.
Second, a governance framework is required to establish the possibility to gain trust in a large variety of issuers. Third, user authenticity must be guaranteed; i.e., sharing and selling credentials must be prevented~\citep{camenisch2001efficient}. However, various concepts that allow one to address these issues, such as initiatives to build governance frameworks~\citep{davie2019trustoverip} and the combination of biometrics, cryptography, economic incentives, and device coupling to create a strong bond between credentials and holders~\citep{hardman2019biometric,hardman2020zkpsavvy}.

In sum, \ac{SSI} allows for highly decentralized management of personal identifiers and for credentials that are reusable across different contexts to be managed by the user from a single app~\citep{sedlmeir2021identities}. In what follows, we investigate how this approach may help meet the challenges of the \ac{eKYC} process.

\section{Method}
\label{sec:methods}
We followed a \ac{DSR} approach. \ac{DSR} was originally created to enable IS practitioners to find solutions to previously unsolved problems through a continual build-and-evaluate process. Its outcomes are IT artifacts, such as constructs, models, methods, or instantiations~\citep{march1995design, hevner2004design}.  While some scholars argue that the IT artifact itself already contributes to research if it is novel and useful~\citep{gregor2013positioning, baskerville2018design}, two challenges in discerning \ac{DSR}'s research contribution remain: First, it is hard to determine what exactly a theoretical contribution in \ac{DSR} is~\citep{gregor2013positioning}. Second, it is hard to balance concrete, practical contributions to a rapidly changing technology environment and to provide a sufficient level of generalization for theory~\citep{baskerville2018design}. To address these challenges, we aim to contribute both an architectural design and a collection of processes as a concrete IT artifact~\citep{gregor2013positioning}. To elevate this IT artifact for further theoretical discussion, we then derive \acp{DP}~\citep{, hevner2004design, gregor2013positioning}. Thus, we aim to contribute nascent design theory in the form of operational principles~\citep{gregor2013positioning}.

For an IT artifact to offer a substantial contribution to IS research, it must address a relevant business need~\citep{hevner2004design}, which can result from the persons, organizations, or technologies used in an environment. As argued in Section~\ref{subsec:KYC}, the enhancement of the \ac{KYC} process represents such a business need. However, an IT artifact must also be applicable in the corresponding environment~\citep{hevner2004design}. To ensure rigor in the design process, the construction of the IT artifact needs to build on existing foundations from previous IS research~\citep{vom2020special}. Also, existing methodologies should be used to evaluate the created artifact~\citep{hevner2004design}. The \ac{KYC} framework here is based on related work that aims to improve the \ac{KYC} process using digital technologies, the technical and theoretical foundations of \ac{KYC}, \ac{DLT}, and \ac{SSI}, and the requirements and expertise of practitioners in said areas.

\begin{figure}[!tb]
    \centering
    \includegraphics[width=\linewidth]{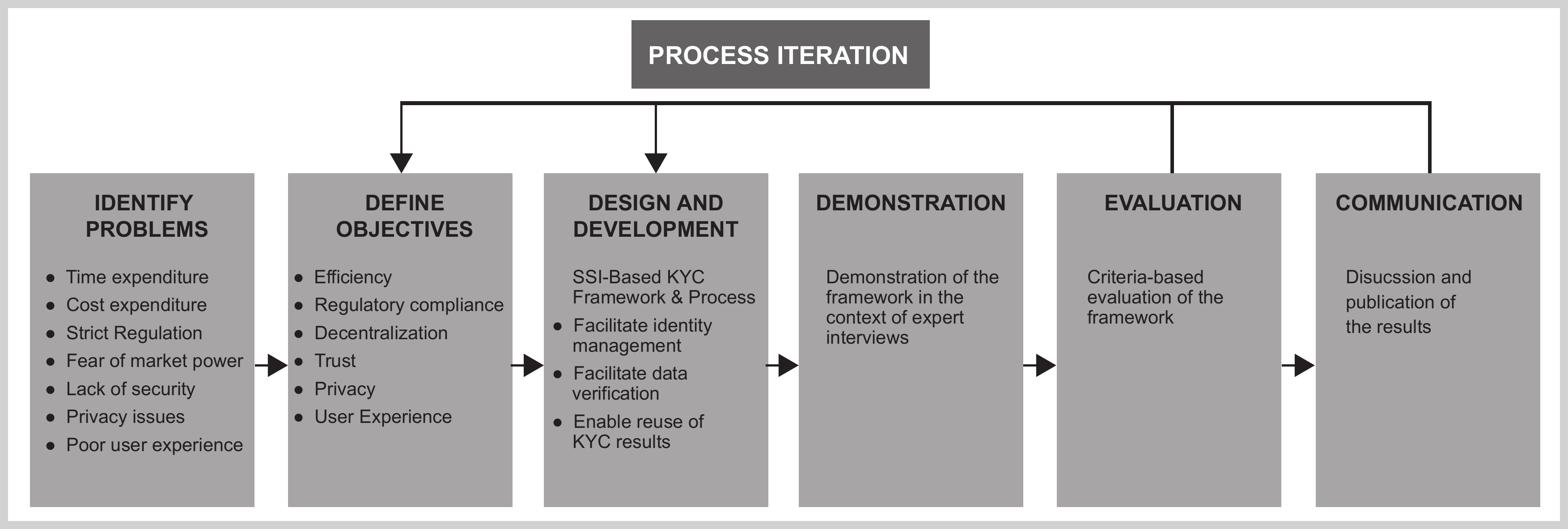}
    \caption{Our applied \ac{DSR} process, following~\cite{peffers2007design}.}
    \label{fig:dsr_process}
\end{figure}

We employ the frequently used and widely accepted~\citep{schweizer2017unchaining,reinecke2013knowing} \ac{DSR} process model of~\citet{peffers2007design} to facilitate the development of a relevant IT artifact created by a rigorous method. Our process has six steps arranged in sequential order (see Figure~\ref{fig:dsr_process}) and incorporates an iterative research procedure by design~\citep{peffers2007design}.
The process typically starts with the identification of a research problem with practical relevance. Indeed, as illustrated in Section~\ref{subsec:KYC}, our examination of the current \ac{KYC} process reveals challenges such as low process efficiency, security challenges, poor user experience, and data protection concerns.

Next, we defined solution objectives to address the stated challenges and to create a meaningful artifact. In line with \ac{DSR}, the insights gained from the build-and-evaluate process must be generalizable and therefore applicable in more generic settings~\citep{jones2007anatomy}. Also, the design artifacts should result in profound disruptions to traditional ways of doing business~\citep{hevner2020envisioning}. Recent research into \ac{DSR} encourages researchers to build their work on prior \ac{DSR} within the respective domain~\citep{vom2020special}. We derived solution objectives by studying the related literature and regulatory requirements, both for the \ac{KYC} process and for digital identification and authentication, resulting in six main objectives for the \ac{KYC} framework and several requirements for each main objective.
Based on these objectives and on theory, we design and develop an \ac{SSI}-based \ac{eKYC} framework in the next research process step. Phase~5 comprises evaluation, which is necessary to test whether an artifact achieves the purpose of its creation and to prove this achievement using rigorous methods~\citep{venable2012comprehensive}. The evaluation phase also helps one to better understand the problem at hand and thus to realize improved outcomes~\citep{hevner2004design}.

There is no unique path regarding evaluation, since the best approach depends on both the underlying problem and the artifact~\citep{peffers2007design}. Our evaluation had several iterative evaluation steps, starting ex ante with the formative evaluation of the design objectives through interviews with experts~\citep{venable2016feds, sonnenberg2012evaluations}. We conducted six additional ex post interviews to summatively evaluate our framework by demonstrating it to the interviewees and incorporating their feedback. The evaluation of the framework was designed to assess its functionality, accuracy, reliability, fit with the organization, and utility~\citep{hevner2004design}. We then applied a criteria-based evaluation concerning whether the derived solution objectives were met, since evaluation criteria for an IT artifact must themselves be determined for the particular environment~\citep{march1995design}. To elevate the implicit knowledge contribution in our IT artifact to more abstract and generalizable knowledge allowing for theoretical discussion~\citep{gregor2013positioning}, we then developed nascent \acp{DP} for blockchain-based SSI, as this technical approach is both novel and increasingly discussed, though no general \acp{DP} currently exist in the literature. Finally, we shared the findings of our research with the relevant audience~\citep{hevner2004design}. The applied \ac{DSR} process was iterative and partly in parallel, since the evaluation phase's results have reshaped the created artifact~\citep{beck2013theory}.

Qualitative interviews, as used for our evaluation cycles, are a frequently used method in IS research, since they are suitable for generating rich data~\citep{myers2007qualitative}. We conducted semi-structured interviews so that we could react flexibly to the interviewees' answers and ask appropriate follow-up questions~\citep{kallio2016systematic}).
We involved experts on \ac{KYC} and \ac{SSI} to reflect opinions from the perspective of practical applicability in existing settings and bank structures as well as opinions regarding technical maturity and feasibility. Also, we took care to avoid an elite bias by representing the voices of executives of different corporate levels~\citep{myers2007qualitative}. Further criteria for the selection of the experts included ample knowledge of their disciplines and intensive experience in their daily work, as well as the ability to provide detailed information on their field of expertise~\citep{morse1991strategies}. A detailed overview over the interviewees appears in Table~\ref{tab:experts}.
\begin{table}
    \centering
    \resizebox{0.70\columnwidth}{!}{
    \begin{tabular}{l|l|l|l|l|l}
\textbf{Episode} & \textbf{Expertise} & \textbf{Id} & \textbf{Role} & \textbf{Background} & \textbf{Type} \\\midrule
1 & KYC & A & Project Manager \ac{KYC} & Strategic Analysis and Research, Banking & Phone Call \\
1 & KYC	& B & Sales Director & Building Society, Banking & Video Call \\
1 & SSI	& C & Identity Engineer & Innovation Consultant & Video Call \\
2 & SSI \& KYC & D & Executive Director & SSI Start-up Founder, Banking & Video Call \\
2 & SSI \& KYC & E & Project Manager & Banking & Phone Call \\
2 & SSI & F & Senior Developer & Computer Science & Video Call \\
2 & SSI	& G & CEO & SSI Start-up Founder & Video Call \\
2 & KYC & H & Sales Executive & Banking & Video call \\
2 & KYC & I & Sales Director & Banking & Video call
    \end{tabular}
    }
    \caption{Overview over the Interviewed Experts.}
    \label{tab:experts}
\end{table}

We recorded 320 interview minutes (an average of 35.6~minutes per interview). The interviews were recorded, transcribed, and later analyzed using MAXQDA~software. For data analysis, we used both open and axial coding~\citep{saldana2015coding}. Starting from the initial concepts originally derived in the open coding round, categories were formed. Categories are ``higher-level concepts under which analysts group lower-level concepts that then become its subcategories''~\citep[p. 220]{corbin2015basics}. During this first coding round, we created 30 categories and 300 subcategories; in the second, we used axial coding to build subcategories. Thus, the data that were split up during open coding were reassembled to summarize the categories on a more abstract level~\citep{saldana2015coding, corbin2015basics, charmaz2006constructing}.

\section{Design Objectives for the eKYC Framework}
\label{sec:objectives}

\subsection{Structuring of design objectives}

To comprehensively address the challenges of the \ac{KYC} process (as identified in Section~\ref{sec:background}), stage~2 in our \ac{DSR} process involved the derivation of objectives to be met by a useful \ac{SSI}-based \ac{eKYC} framework. We derived these objectives from  the literature on the \ac{KYC} process, \ac{KYC}-related regulatory requirements, and three formative interviews with experts. Thus, we aimed to align with \ac{DSR} by incorporating prior research~\citep{vom2020special} and incorporating real-world business needs~\citep{hevner2004design}. We identified six main objectives and associated requirements. In what follows, we explain and justify them.

\subsubsection*{Objective 1: Efficiency}
The high cost and human resources involved in carrying out the current \ac{KYC} process strongly challenges financial institutions. Financial institutions offering fast and convenient verification of identity documents are more attractive from the customer's perspective, can reduce costs, and can help a company gain a competitive advantage~\citep{jessel2018digital}. To allow for increased process efficiency, we derived three requirements that had to be satisfied to overcome the existing challenges. The \emph{end-to-end digital processing of relevant documents~(R\,1.1)} is a prerequisite for automating process steps and reducing friction~\citep{arner2019identity}. Further, in the current \ac{KYC} process, many steps involving the validation of data, such as checking whether an ID document's validity has expired, are conducted manually~\citep{zetzsche2018digital}. Thus, the de facto \emph{automation of manual processes~(R\,1.2)} is another key requirement. Further, \citeauthor{moyano2017kyc}'s~[\citeyear{moyano2017kyc}] interviews with five senior banking executives revealed a need for interbank collaboration; this was confirmed by Experts~A and~B, who stated that banks would be ready to collaborate on resource-intensive KYC.  Currently, however, the main barrier to such cross-organizational processes is the lack of a suitable non-proprietary IT infrastructure. Thus, a \emph{standardized exchange of \ac{eKYC} documents~(R\,1.3)} is crucial to allow for the efficient integration of \ac{eKYC} checks that have been conducted at other institutions.

\subsubsection*{Objective 2: Regulatory compliance}
Compliance with regulations is a key objective of the \ac{KYC} process \citep{Ostern2020blockchain}; derived from the overall goal of avoiding money laundering, it is one of the main reasons why the \ac{KYC} process exists at all.
Our literature study revealed that the \emph{\ac{MLA} (R\,2.1)}, \emph{\ac{GDPR} (R\,2.2)}, and \emph{\ac{eIDAS} (R\,2.3)} are particularly relevant regulations for a digital \ac{KYC} process~\citep{arner2019identity}. While these apply within the \ac{EU}, there are similar regulatory requirements in other jurisdictions worldwide. The European requirements are considered particularly strict, which is why we decided to apply them here.

The \ac{MLA} provides banks with specific requirements regarding the identification of customers and the storage of their records. The banks are also required to determine and document the risk in relation to their customers. The \ac{GDPR} applies to the processing of any data regarding natural persons, but not to legal entities, and poses requirements such as privacy by design, portability, the right to erasure, transparency, purpose limitation, data minimization, accuracy, storage limitation, information integrity, and confidentiality. Further, digital \ac{KYC} processes involve the customer's identification and a check of the authenticity of the involved documents, and the 5th \ac{EU} \ac{AML} Directive accepts electronic ID systems that comply with \ac{eIDAS} as a legitimate means of identification for \ac{KYC} procedures. \ac{eIDAS} imposes requirements on these electronic means of identification, such as compliance with certain security levels (level of assurance) and the cross-border interoperability of systems.

\subsubsection*{Objective 3: Decentralization}
As argued in Section~\ref{subsec:KYC}, silos of customer data are an attractive target for attackers. Securing valuable information is costly and not the core business of banks, and mistakes can have severe consequences concerning reputation, fines, or both. Recent data breaches that revealed sensitive customer data stored in central data silos have significantly reduced confidence in their respective architectures~\citep{rajput2017towards}. To avoid comparable data breaches, a viable solution for an improved \ac{eKYC} process must therefore \emph{avoid central storage of customer data (R\,3.1)}. Further, banks do not want to risk becoming dependent on a centralized \ac{eKYC} service provider. Thus, the system must be constructed to \emph{prevent lock-in effects (R\,3.2)} that could result in the aggregation of market power. Decentralization of both data storage and workflows is therefore one key objective of the new \ac{eKYC} architecture.

\subsubsection*{Objective 4: Trust}
A key goal of banks is to make \ac{eKYC} documents reusable in registrations of a customer at different banks. If banks do not comply with the regulations, there can be heavy fines, so it is important to establish trust in the \ac{KYC} process and the integrity of its documentation at other banks.
Thus, \emph{acceptance of \ac{KYC} documents attested by other banks (R\,4.1)} is required. The documents must be tamper-proof, so a further requirement is that \emph{validity checks (R\,4.2)} of these documents are feasible.
Another often disregarded requirement for a complete trust chain is that sharing or selling \ac{KYC} documents among customers must be prevented. This can be particularly difficult if the \ac{eKYC} process happens remotely and lacks interaction with an employee of the bank. The customer needs to be able to convince the bank that the \ac{KYC}-related documents that they present were not stolen, sold, or shared. We call this requirement \emph{authenticity checks (R\,4.3)}, meaning that the identity of the customer and their connection with the documents must have a high level of assurance even if the customer is not present at a branch and no video call is held.

\subsubsection*{Objective 5: Privacy}
Protecting customers' privacy is a key feature of an \ac{eKYC} process. Facing an increasing number of data leaks, customers are aware of privacy issues, and delivering a privacy preserving solution may increase the solution's acceptance.
An essential and fairly universal principle in this context is \emph{compliance with the ``need to know'' principle (R\,5.1)}: Only the customers themselves and entities relevant to the \ac{KYC} process must have access to customers' personal data. This is also a general recommendation for information systems from a security perspective~\citep{hughes1988need, moor1997towards}.
Further, not only the parties involved in the \ac{KYC} process but also the de facto data that are exchanged should be restricted to what is necessary, because digital data are much more comprehensive and easier to collect and abuse than their analog counterparts~\citep{arner2019identity}. We call this requirement \emph{data minimization (R\,5.2)}.

\subsubsection*{Objective 6: User experience}
From the users' perspective, although privacy is a nice feature that can be used for marketing purposes, the most important objective is seamless user experience~\citep{kokolakis2017privacy}. The \ac{eKYC} process must be convenient, so that customers are not discouraged from registering at the new bank. It is only when the \ac{eKYC} process is fast and simple for the customer that it can provide high security and acceptance~\citep{dhamija2008seven}. Thus, we made \emph{low complexity (R\,6.1)} a major requirement for user experience. Further, the variety of devices on which a customer can perform the \ac{eKYC} process must be respected. Mobile phones are often the customers' preferred option, but support for web apps is also necessary in many circumstances. Thus, the availability of \emph{different user interfaces~(R\,6.2)} is important. The user experience should also include exception handling, for instance, if a device that stores the customer data is lost or stolen. In this case, either there must be a built-in recovery mechanism, or the customer must be able to ask for rapid support. This is very difficult if no central third party is responsible for the whole process. Thus, we also added such \emph{backup, recovery, and support (R\,6.3)} features to our requirements.

\subsection{Evaluation of the design objectives}
\label{subsec:evaluation}
We discussed the current problems of the \ac{KYC} procedure and our derived objectives with two \ac{KYC} experts and an \ac{SSI} expert.
The interviews sought to evaluate the identified design objectives concerning relevance and completeness. The \ac{KYC} experts worked in different companies and held different positions, so that the objectives could be viewed from different perspectives. Additional information on the interviewees appears in Table~\ref{tab:experts}.

Expert~A confirmed the relevance of the derived objectives and their associated requirements. Owing to the increasing expenditure on personnel and technology, the process's efficiency is indeed a crucial goal for banks. He stressed the importance of end-to-end digital processing and advocated interbank cooperation in the \ac{KYC} process, but identified trust problems here, both between the banks and concerning customer trust in the confidentiality of their data. According to him, the protection of customer privacy is also crucial. Further, he affirmed the relevance of increasingly strict regulations and the need to comply with them. For instance, customer data must be stored by banks for at least five years. The expert also confirmed the necessity of including further \ac{MLA} requirements.

Expert~B also described process efficiency as the most crucial factor, to ensure cost and time savings. The challenges apparently lie particularly in the high number of manual process steps. This expert emphasized the importance of automation and digital processing of documents. He also confirmed the importance of protecting privacy. Sensitive handling of customer data is necessary, and this must not be passed on to third parties, not even to cooperation partners. Like Expert~A, he noted the increasing importance of regulation and the need to comply with it.

Expert~C emphasized the importance of a good user experience, since many users will not focus on the systems' functional details. During the implementation phase, special care should be taken to ensure that the system is as intuitive as possible. Asked about the architectural perspective, he mentioned backup and recovery capabilities through cloud storage as a building block for user friendliness in case of data theft or loss.
Expert~C also confirmed the importance of the \ac{GDPR} and \ac{eIDAS}. According to him, there is still room for interpretation in the \ac{GDPR}, for instance regarding the role of encrypted or hashed personally identifying data. He advised proceeding from the strictest possible interpretation of the \ac{GDPR}. He stressed that, on a distributed ledger, data cannot be deleted. A key challenge to the acceptance of \ac{KYC} documents attested by other banks, he spoke of the necessary establishment of a trust relationship between the banks. However, he argued that connecting the \ac{eKYC} architecture to the \ac{eIDAS} infrastructure could be a solution to this problem.

In sum, at least one expert emphasized each of the design objectives, and the experts generally considered the associated requirements useful to evaluate an \ac{eKYC} framework from a bank's perspective.

\section{A Framework for eKYC Processes Built on Blockchain-Based SSI}
\label{sec:framework}

\subsection{The SSI-based eKYC architecture}
Based on the related work presented in Section~\ref{sec:background}, we designed a decentralized architecture that seeks to address the challenges of the KYC process. The study by \citep{moyano2017kyc} motivated a decentralized design of \ac{eKYC} to allow for inter-bank collaboration. However, the proposed system seems critical from a data protection perspective, considering the privacy-related challenges of storing customer data, also in encrypted or hashed form, on a blockchain system. We also noticed that the mechanism they presented does not enforce the alignment of incentives, since the integrity check only requires a local read operation on one node. Thus, while we appreciate the background they gave on the necessity of a reusable \ac{KYC} and a non-centralized solution, we found the design of the \ac{SSI}-based framework proposed by~\citep{soltani2018new} more appropriate. Nonetheless, we generalized this solution by considering both the initial onboarding to receive the first \ac{KYC} document and how an existing \ac{SSI} ecosystem and regulation such as \ac{eIDAS} integrates with and further strengthens \ac{eKYC}. We also added an investigation of the framework's practical feasibility by rigorously evaluating our \ac{SSI}-based framework concerning the technical, economic, and legal requirements. From a technical perspective, we abstracted from their solution based on Hyperledger Indy to a more generic perspective on \ac{SSI} and extended their findings by rigorously evaluating the design.

\begin{figure*}[!tb]
    \centering
    \includegraphics[width=\linewidth]{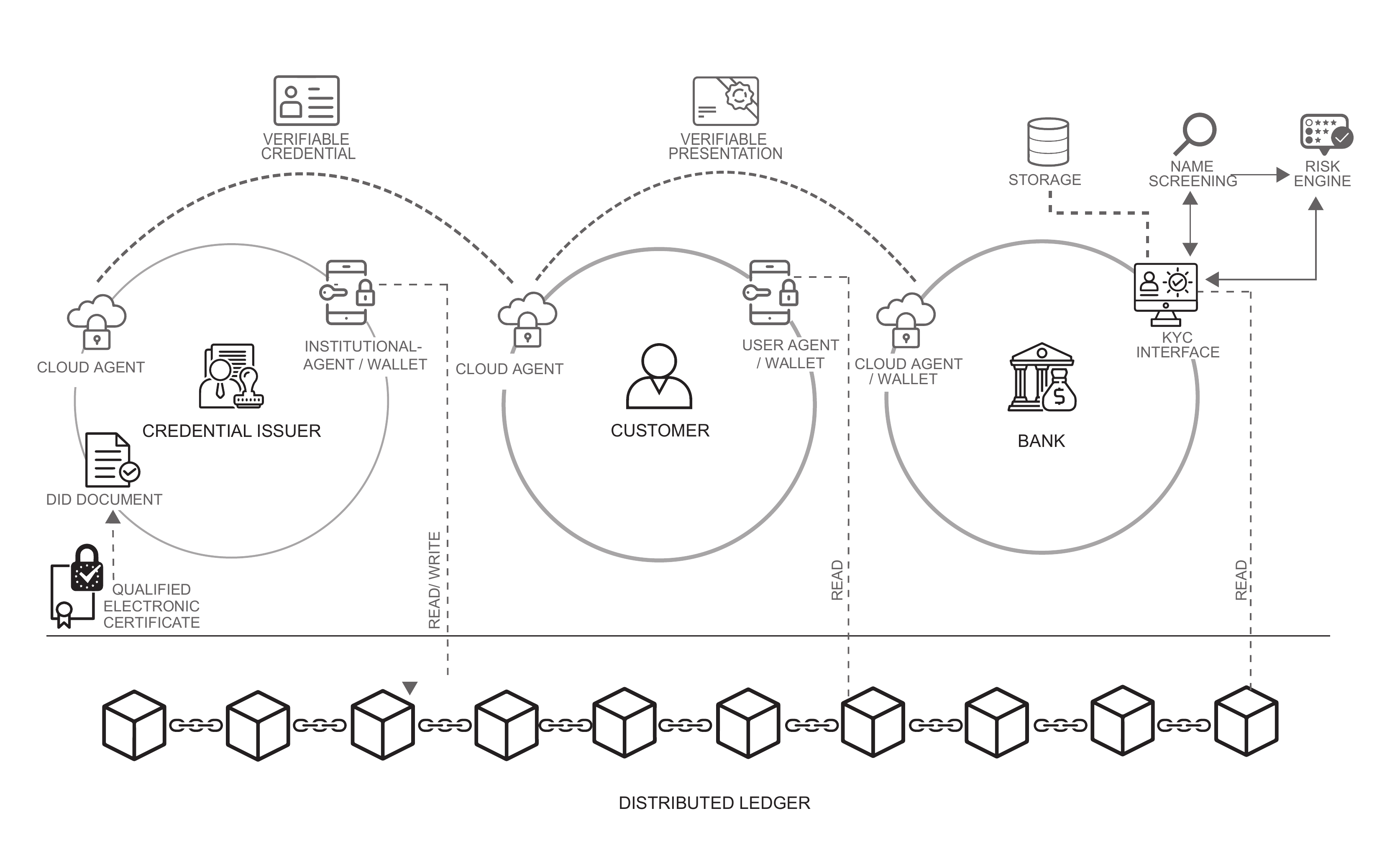}
    \caption{\ac{SSI}-based \ac{KYC} architecture (based on~\citep{eu2019ssieidas,moyano2017kyc,reed2018dkms,soltani2018new,w3c2020did,w3c2019vc}).}
    \label{fig:KYC_refined_architecture}
\end{figure*}

Following~\citep{soltani2018new} and the general approach of blockchain-based \ac{SSI}, the proposed \ac{eKYC} architecture involves three primary parties: the customer (holder), a bank (verifier), and an issuer (the same bank, another bank, or any third party trusted by the verifying bank, such as a government agency). Credentials from different issuers can also be conjugated, because in a so-called \ac{VP}, attributes attested in different \acp{VC} can be combined~\citep{w3c2019vc}. For simplicity, we assumed one issuer. The customer is the \ac{KYC} subject and defines the center of the architecture (see Figure~\ref{fig:KYC_refined_architecture}). Customers manage their digital identity through user agents by creating and storing \acp{DID} and cryptographic keys in their digital wallets, collecting credentials, creating backups, and managing permissions. It is possible to interact with agents on various devices, such as smartphones or laptops. At all times, the customers have full control over their data and particularly over \ac{KYC}-related documents, represented by \acp{VC}. While traditional certificate-based approaches (e.g., X.509 certificates) need to be shown fully to the verifier in order to check the signature's validity, the \ac{VC} standard~\citep{w3c2019vc} and related implementations such as Hyperledger Aries allow for creating proofs from the \acp{VC}, convincing the verifier that certain claims extracted from the \ac{VC} are correct without the need to exchange the full \ac{VC}. This builds on cryptographic constructions such as anonymous credentials introduced by \citep{camenisch2001efficient}.
In our case, the \acp{VP} contain proofs of the validity of the attributes that need to be revealed during the \ac{KYC} process.

To facilitate the redundant storage of credentials and easier user access to the \ac{SSI} documents, as well as to enable secure communication with other entities, the framework employs cloud agents and wallets. The permissions for carrying out identification activities differ between edge and cloud instances. While edge agents and wallets are usually granted full access to an individual's data, the user should use cloud agents/wallets primarily for redundant storage and communication with other entities. A blockchain serves as a neutral infrastructure for storing publicly verifiable information. It is used to hold \ac{VC} issuers' public signing keys and other institutional information. Further, schemas of \ac{KYC} \acp{VC} are stored on-chain to allow for public verification. Also, publicly available revocation registries are stored on a blockchain to allow for public checks of privacy-preserving revocation information. Which credentials are accepted in the \ac{KYC} process may be defined by each bank, depending on its requirements and trust relationships. The combination of \ac{eIDAS} and \acp{DID} could allow for qualified digital signatures that comply with \ac{eIDAS}~\citep{eu2019ssieidas}. Credential issuers use institutional agents that are explicitly designed for creating credentials.
Besides issuing credentials, these agents perform identification activities such as checking credentials for integrity and direct communication with the customer that is relevant during and after the \ac{KYC} process. It also has an interface to name screening services, the bank's risk engine, and customer monitoring. The financial institutions are obliged to store data about customers, for which they use separate storage.

\subsection{The SSI-based eKYC process}
In accordance with the generic procedure of \ac{KYC} processes, we split the proposed \ac{SSI}-based \ac{eKYC} process into three parts: (1) customer identification, data verification, and identity authentication; (2) name screening, risk assessment, and enhanced due diligence; and (3) ongoing monitoring and records keeping. The first part involves three scenarios, depending on the customer's status in the \ac{KYC} process.

\subsubsection*{Customer identification, data verification, and identity authentication}

The \ac{KYC} process starts with customer onboarding, where three cases can be distinguished. The first case is \emph{completely new onboarding}, where the customer has neither an \ac{SSI} agent/wallet nor \acp{VC} that confirm a completed \ac{KYC} process. The second case, \emph{fast onboarding}, is possible if the customer already has an \ac{SSI} agent/wallet with corresponding \acp{VC} that attest to the prior completion of a \ac{KYC} process. Third, we discuss a simplified case we call \emph{new to \ac{KYC}}, where customers already have an \ac{SSI} agent or wallet and some \ac{VC} from other contexts that contain identity-related information trusted by the verifying bank (or that the bank is allowed to trust from a legal perspective), but do not yet have \acp{VC} that demonstrate the completion of the \ac{KYC} process at some institution.

\begin{figure}[!tb]
    \centering
    \includegraphics[width=\linewidth]{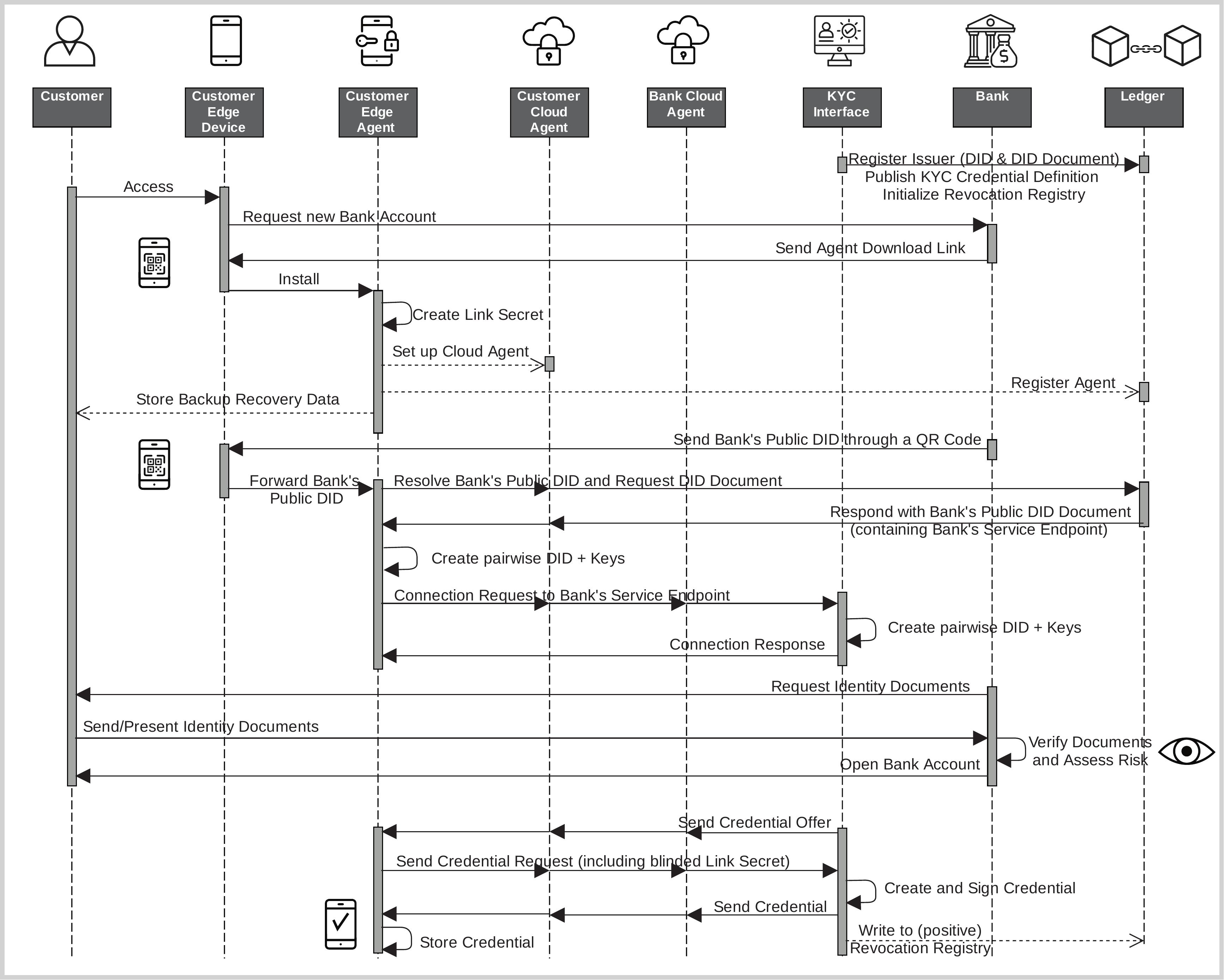}
    \caption{UML Diagram: \emph{Completely New Onboarding}}
    \label{fig:kyc_cno}
\end{figure}

We present the first case, \emph{completely new onboarding}, in a UML~sequence diagram (see Figure~\ref{fig:kyc_cno}). To enable \ac{SSI}-based onboarding, as illustrated in \citep{soltani2018new}, banks must conduct a one-time bootstrapping process in which they first store a public \ac{DID} and an associated \ac{DID} document in a distributed ledger. This \ac{DID} document may contain service endpoints of the bank, e.g., for obtaining customer services or conducting the \ac{eKYC}. The bank will also publish a so-called credential definition, which may be derived from an agreed-on schema/template that contains the attributes that should be attested in a credential, and a revocation registry. All this information is meant to be publicly readable and contains cryptographic information that allows banks (verifiers) to check the validity of \acp{VP} that use an associated \ac{VC}, and customers (holders) to conduct proofs of non-revocation. Thus, these can be stored on the blockchain layer, and no \ac{GDPR}-related problems are to be expected.

After this initial setup, the bank is ready to perform customer onboarding processes. While \citep{soltani2018new} presented a (slightly less detailed) sequence diagram for customer onboarding, it involves reading from and writing to the blockchain more often than technically necessary. According to our interview with a co-author of the \ac{W3C} \ac{DID} standard, it suffices and is preferable from a privacy perspective to have a peer \ac{DID} for the customer. We also discuss in detail the implications of the design for the bank and the customer, for instance, related to binding, revocation, and backups.
\emph{Completely new onboarding} starts with a bank customer who either visits the bank's website with their smartphone or laptop or physically arrives at a local bank branch to open a new bank account. Since the customer has neither an \ac{SSI} user wallet nor the necessary \ac{KYC} credentials, the bank recommends or offers a user wallet and provides the customer with a corresponding download link.
Customers can download any digital wallet of their choice that supports the  public \ac{DID}, peer DID, and \ac{VC} standards. It stores credentials and keys and is secured by a password or biometrics. The bank could further offer an edge agent in encrypted form in the cloud as capability for backup and recovery. The user wallet creates a new \ac{DID} and some associated keys required for encryption and stores them in the wallet. At the start, the user also creates a so-called link secret, which will later be used to tie different credentials together in a \ac{VC} and thus provides a means to prevent selective credential sharing. However, as long as all credentials contain the customer's name or another strongly binding attribute that needs to be revealed in the \ac{VP}, it is easy to prove that they belong together also without a link secret.

The customer can now use their newly generated \ac{DID} to establish an end-to-end-encrypted (secure) connection to the endpoint that the bank offers for the \ac{eKYC}. The bank could also provide this information by submitting a QR~code to the customer (e.g., via e-mail). The customer scans this QR~code with their wallet app and thus connects to the bank's public service endpoint. The bank's service behind this endpoint now also creates a new pairwise \ac{DID} as well as a key pair that the bank will use exclusively in this relationship, and sends a connection request to the customer's service endpoint, its cloud agent, which forwards the connection request to the customer's wallet app. This connection request contains the bank's pairwise \ac{DID}, the public key used by the bank, and the service endpoint at which the customer can contact the bank, and could also involve a proof that the pairwise \ac{DID} has in fact been authorized by the bank (e.g., through a \ac{VP} in which the bank reveals its legal identifier that has been certified by a reputable public institution). In turn, the customer's digital wallet checks the connection's authenticity and creates a pairwise \ac{DID} and keys for the relationship with the bank. Next, it sends a connection response to the bank's cloud agent/wallet, which forwards it to the bank's \ac{KYC} interface. Now an end-to-end-encrypted connection exists between the bank and the customer, which can be used to securely exchange messages, public keys, \acp{VC}, and \acp{VP}.
Since the customer does not yet have \acp{VC}, the customer's identity must first be verified. The customer sends the necessary analog identity data to the bank, either by traditional means or -- if feasible -- in scanned form via e-mail or the just-established connection. If the customer opens an account in a bank branch, the documents can also be verified directly there.

After the data have been verified and the customer's identity has been confirmed, the bank can send a credential offer to the customer's edge user agent via the established connection. This credential offer contains a preview of the data that will be attested, the credential issuer information, an expiration date for the \ac{VC}, and information regarding credential revocation.
The customer then accepts the credential offer and sends it to the bank, containing the link secret in blinded form.\footnote{To be precise, the blinded link secret is a cryptographic commitment, i.e., the hash of the link secret and some one-time random number. Thus, while the blinded form will differ in each credential issuing process, the customer (holder) can still prove that different commitments originate from the same link secret, without revealing the link secret itself, in a \ac{ZKP}.} However, the customer only has to create the link secret once and can later reuse it for their other \acp{VC}. The bank includes the blinded link secret in the attributes attested in the \ac{VC} and sends the \ac{VC} to the customer. The credential could support selective disclosure. That is, in any \ac{VP}, the customer can include only the attributes attested by the \ac{VC} that are necessary for the verifier, and combine claims from different \acp{VC} into a \ac{VP}.

If the issuer wants to support revocation and has bootstrapped a revocation registry, the \ac{VC} also contains information on how to check its revocation status. The credential issuer can then revoke credentials by updating a revocation registry in the distributed ledger. The bank, for instance, can use this mechanism to invalidate a credential that turns out to be wrongly issued. Notably, it is only through the additional information regarding revocation in the credential that the customer can make sense of the information in the public revocation registry and create a proof of non-revocation within a \ac{VP} that contains attributes from this \ac{VC}. Since the credential is never revealed, but only proofs are derived from this, this likely makes public revocation registries compliant with the \ac{GDPR}.

Going beyond~\citep{soltani2018new}, we present in detail how the reuse of a \ac{KYC} process works. This \emph{fast onboarding} process also begins with a bank customer visiting the bank's website or a local bank branch to open a new bank account. The customer states that they already have an \ac{SSI} user agent and \acp{VC}. In the case of opening an account online, the bank sends the customer its bank public \ac{DID}, for instance by means of a QR~code that can be scanned by the customer's wallet app. The channel by which the customer receives this information must be trusted, as the customer does not know the bank's \ac{DID} in advance. Using an identity infrastructure such as \ac{eIDAS}, the customer could check whether they are really communicating with the corresponding bank by verifying an \ac{eIDAS} certification on the bank's public key in the \ac{DID} document. The customer's user agent can then identify the distributed ledger that stores the \ac{DID} document associated with the \ac{DID} and can query this ledger for the \ac{DID} document. The user agent uses the \ac{DID} document to identify the bank's \ac{eKYC}-related service endpoint. The customer's user agent now creates a pairwise \ac{DID} for this relationship and the corresponding keys and sends a connection request to the bank. The connection request also contains the customer's pairwise \ac{DID} and the public key used. The bank then creates a pairwise \ac{DID} and corresponding keys, and sends the customer a connection response, including the pairwise \ac{DID} and public key.

\begin{figure}[!htb]
    \centering
    \includegraphics[width=\linewidth]{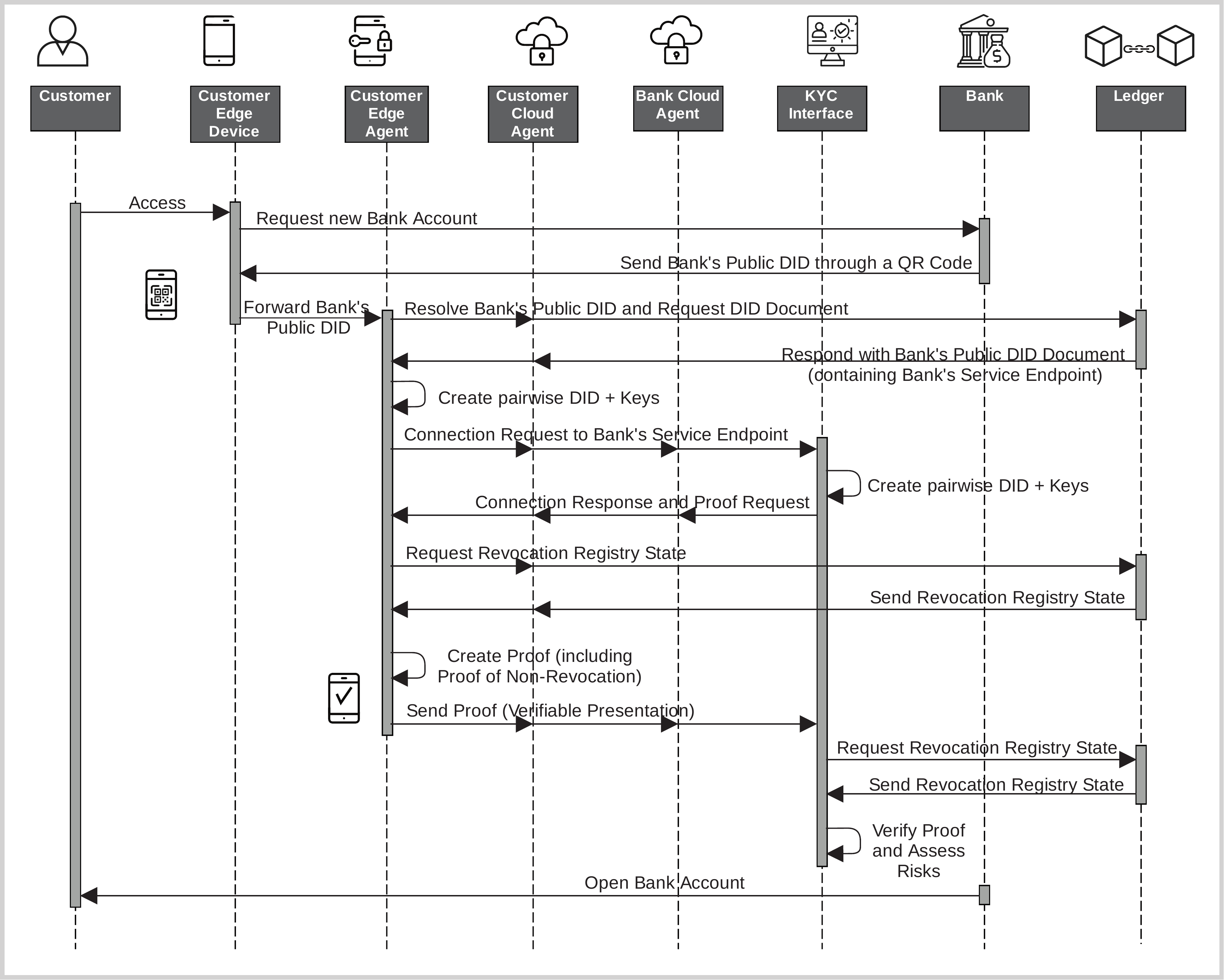}
    \caption{UML Diagram: \emph{Fast Onboarding}}
    \label{fig:kyc_fo}
\end{figure}

After establishing the secure connection, the bank sends a proof request for conducting \emph{fast onboarding} \ac{KYC}. This request contains a random nonce to prevent replay attacks and specifies which data the customer must transmit to the bank, and restrictions on when to accept the \ac{VP}. This includes a specification of issuers (credential definitions) and schemas that are accepted for the \acp{VC} used, and whether there is a need for a proof of non-revocation, including a timestamp of the revocation registry that the customer should refer to in creating this proof if a proof of non-revocation is demanded. The customer's edge agent automatically searches for \acp{VC} stored in the customer's digital wallet that match these requirements, updates their local revocation registry through a query if it has not been cached before, and creates a \ac{VP} that it sends to the bank. The bank can now cryptographically verify the claims, which may also involve reading from revocation registries and other information regarding credential definitions unless sufficiently timely local data from previous queries are cached.

The bank can now cryptographically verify the proof, which involves checks that the digital signatures of issuers were on all attributes involved in the \ac{VP}, that none of the attested attributes came from a revoked \ac{VC}, and that all the \acp{VC} involved were issued to a commitment of the same, common link secret. After the proof has been verified, the bank account is opened. This whole process can be highly automated and is completed in a few seconds. The secure channel established between the bank and the customer based on pairwise \acp{DID} can be used in the future to exchange further documents and to communicate securely and reliably. Based on the exchanged keys, a unique authentication of the involved entities is thereby possible. This is important when dealing with digital identities, since the bank must ensure that it communicates with the same person over time~\citep{jessel2018digital}.

The third case is \emph{new to \ac{KYC}}, where the customer already has an \ac{SSI} user agent and maybe even identity-related \acp{VC}, but does not yet have a \ac{VC} that is accepted by the banks during the \ac{KYC} process. Thus, \emph{new to KYC} is a combination of \emph{completely new onboarding} (Figure~\ref{fig:kyc_cno}) and \emph{fast onboarding}~(Figure~\ref{fig:kyc_fo}). While the construction of the pairwise \ac{DID} relationship between the bank and the customer corresponds to the fast onboarding process, the transmission of the analog ID documents and the possibility of getting a \ac{KYC} credential from the bank corresponds to the process of \emph{completely new onboarding}. However, the customer could first check, through a proof request, whether only a subset of ID documents is necessary because some digital identity proofs are already in their wallet. In addition to the option in which the customer opens an account online, it is also possible to open an account directly in a bank branch by using a QR~code to receive the bank's service endpoint and have the analog ID documents checked directly in the bank.

\subsubsection*{Name screening, risk assessment, and enhanced due diligence}
After the identity data has been exchanged and cryptographically verified, the name screening service runs in the background of the bank's IT system to check the data against publicly known blacklists regarding terrorism, illegal money laundering activities, politically exposed persons, and negative press. The result of the name screening service is then fed directly into the risk engine, which uses this and other information to classify the customer into a risk class.
The risk engine then calculates a risk score and classifies the customer into low, standard, or high risk. Depending on this result, further checks may be necessary before the bank opens the account. Since, in contrast to analog ID documents, \acp{VC} are much harder to forge, it suffices to request a minimum amount of information at the start of the relationship. Depending on the risk assessment's result, additional checks may become necessary later. To mitigate risk, the bank can use the previously established secure communication channel to request additional documents and information, such as an income statement or the reason for opening the account. As illustrated, such additional documentation could again be provided in analog form or by using \acp{VC} that are already in the customer's wallet -- for instance, an income statement issued by an employer that the verifying bank trusts -- and deliver an associated \ac{VP}. Once the customer's verification is successfully completed, the bank can open the account.

\subsubsection*{Ongoing monitoring}
Once the account has been opened, the risk engine checks the customer's ongoing transactions during the business relationship, compares these to the expected transaction volume, and checks the transactions for suspicious transaction patterns. Further, the risk engine regularly checks whether the expiration dates contained in the \acp{VP} have expired; these may even occasionally trigger a new proof request to the customer through the secure connection to ensure that none of the customer's \acp{VC} that were used for \ac{KYC} have been revoked. The customer then only has to press a confirm button to deliver a new \ac{VP}. Thus, a manual check of the identity documents is no longer necessary. If it turns out in a periodic refreshment that a customer's \ac{VC} has in the meantime been revoked (which may just be because of a change of address or a successive re-issuance of an ID card) or that the transaction behavior is abnormal, the risk engine reassesses the risk and proposes measures to mitigate these risks if necessary. The bank can then also request an updated version of the customer's \acp{VC} or further documents. This could even be extended to offering the customer an option to automatically send updated versions of their \acp{VC} (e.g., if the address on the customer's government-issued identity \ac{VC} changes) to the bank after the \ac{KYC} process, so that no more manual activities by the bank and the customer are necessary to keep the data up to date.

\subsubsection*{Record keeping}
The \ac{SSI} concept theoretically allows a bank not to store personal data about its customers at all. The data are solely stored in the customer's digital wallet, and it is very easy to request data when needed and convenient for the customer to provide this information. However, depending on the specific regulations, the banks may be obliged to store their customers' data for a longer period in order to be able to unambiguously determine the person's identity in the event of suspicious or illegal conduct. Therefore, the bank also stores the data in a local database in the redefined \ac{KYC} process. However, the bank may still manipulate some data. A benefit of \ac{ZKP}-oriented \acp{VC} is that the \ac{VP} could be made either repudiable or non-repudiable, ensuring tamper-proof documentation where required and supporting customer privacy even in the case of hacks if no auditability is required or sensitive information such as income is involved~\citep{hardman2020zkpsavvy}.

\section{Evaluation}
\label{sec:evaluation}
We now report on a summative, criteria-based evaluation of our proposed framework with interviews with experts (as described in Section~\ref{sec:methods}), evaluating each of the objectives derived in Section~\ref{sec:objectives} and their associated requirements in detail~\citep{march1995design}.

\subsubsection*{Efficiency}

According to Experts~E~and~I, the \ac{SSI}-based KYC process presented in the framework has the potential to solve the inefficiencies in the existing \ac{KYC} process. This can mainly be achieved because the framework involves fully digital cryptographic proofs in the form of \acp{VC}.
By processing the data entirely digitally (R\,1.1), friction in the onboarding process can therefore be reduced for both the customer and the bank (Experts~D,\,H, and I).
The need for face-to-face verification, manual data processing, and repeated \ac{KYC} processes can be eliminated through the use of re-usable  \acp{VC} combined with revocation registries on the blockchain, thus saving costs for manual and repeated process steps (R\,1.2).
Expert~H also emphasized that updates and periodic confirmation that customers need to provide to banks regarding their data can be significantly reduced through the bilateral and secure communication channel, through which the customer can easily give \acp{VP} to the bank. In addition to the potential personnel cost savings, the possibility of authentication with a high level of assurance and the associated reduction of risks can also avoid high penalties for non-compliance with due diligence regulations and standards. However, if there is not yet an existing ecosystem of official identity-related documents, this is only true for the \emph{fast onboarding} process, where a prior \ac{eKYC} process at another bank or official document issuer has taken place. Standards for \ac{KYC} credentials can be created and stored on a public blockchain, such that they can be referred to and accepted by a range of institutions (R\,1.3). This standardization can be particularly valuable when verification of unknown foreign documents can be avoided~(Expert~I). Nonetheless, questions regarding governance (e.g., who defines standards) remain open. An additional governance framework is therefore necessary to create clear guidelines for defining which institutions are suitable as credential issuers.

\subsubsection*{Regulatory compliance}

Regarding compliance with the \ac{MLA}, the interviewees did not see particular difficulties in the framework design (R\,2.1).
The \ac{GDPR} grants the right to erasure of personal data if the reason for their processing no longer exists. While the de facto interpretation of this regulation remains unclear, it must be assumed that encrypted and hashed personal data also fall under this regulation (Expert~C). Further, public \acp{DID} and public keys could be considered as personal data under \ac{GDPR}, and must therefore be deleted if customers request this (Expert~E). Thus, \ac{KYC} designs that use distributed ledgers to store such data cannot be implemented by banks. In our framework, natural persons only use pairwise \acp{DID} and exchange information bilaterally without writing it to a distributed ledger (Experts~E,\,F, and G). Further \ac{GDPR} requirements, such as data minimization, are also naturally addressed through \acp{VC}' selective disclosure capabilities.
The fundamental objectives of our \ac{eKYC} process are therefore aligned with those of the \ac{GDPR}~(R\,2.2). However, a detailed legal assessment remains an avenue for future research.

To effectively use the system, it must also comply with \ac{eIDAS} regulation (Expert~F). The experts noted that they do not see a conflict between \ac{eIDAS} and the \ac{SSI}-based \ac{eKYC} process~(R\,2.3), and
supported the idea of combining the \ac{SSI} concept and the \ac{eIDAS} infrastructure (Experts~D,\,E,\,F,\,G, and I). Expert~G stressed that ``these regulations are drivers that will help to adopt \ac{SSI}, because \ac{SSI} is an ideal way to implement them.'' The \ac{EU} has started building the \emph{\ac{eIDAS} bridge}, which seeks to make the legally qualified signatures from \ac{eIDAS} accessible for the \ac{VC} standard; nonetheless, this implementation has not yet been completed.

\subsubsection*{Decentralization}

Our framework for \ac{eKYC} stores identity-related data in the customer's digital wallet, i.e., on a mobile phone or laptop. Besides the need for banks to store customer information for a certain period -- owing to regulatory compliance, rather than for technical reasons -- central storage is therefore unnecessary (R\,3.1).
User agents, whose role is discussed in R\,6.3, could be considered for centralized storage.
However, these only store data encrypted under user-managed keys. Further, owing to the heavy standardization associated with \ac{SSI}, it is unlikely that user agents hosted by third parties will encounter the same network effects that have led to centralization for traditional identity providers in federated systems. Thus, the framework counteracts data silos that are highly attractive to hackers, since one can no longer capture many data sets at once (Experts~E and F).

The proposed framework also induces no new central parties to the \ac{KYC} process (R\,3.2). Through the use of blockchain, no single entity controls the infrastructure that is involved in checking credential schemas or revocation registries.
Expert~F mentioned the banks' position of trust toward their customers, and therefore considered the banks to be very suitable providers of cloud agents and wallets. Further, most of the experts support the idea of using banks as potential service providers to backup facilities (Experts~E,\,F, and G).

\subsubsection*{Trust}

In our framework, \acp{VC} form the basis of \ac{KYC} documents. The combination of \acp{VC}, as an evolution of digital certificates with additional capabilities such as selective disclosure and privacy-preserving revocation mechanisms based on a blockchain, yield a natural digital equivalent of physical \ac{KYC} documents that customers can fully control and take to other banks. \acp{VC}' integrity can be tested by checking the digital signatures' validity, whereby the signing keys of issuing institutions (such as other banks) are publicly available on a blockchain. It is not possible to create fake credentials, because these are not valid without a credential issuer's signature. Further, ownership of credentials can be cryptographically proven~\citep{rannenberg2015attribute}, and binding multiple credentials via a strongly correlating attribute such as the holder's name or biometric properties, or cryptographically through secure hardware, makes credential sharing or selling difficult and unattractive (R\,4.3). From the perspective of both banks and regulators, fully digital verification provides a significant advantage over analog documents, since data accuracy is improved and manual errors during data processing can be ruled out (R\,4.1) (Experts~D,\,E, and I).

The use of the blockchain infrastructure for storing information on credential issuers (e.g., other banks or government institutions) and revocation registries for \acp{VC} provides an infrastructure that allows a bank to verify \acp{VC} issued by other banks. Nonetheless, governance mechanisms regarding the legal acceptance of such \acp{VC} and other aspects of interbank collaboration required for (R\,4.2) still leave some questions open (Experts~D,\,F,\,G, and I). Such a framework is necessary to clarify which credentials the banks accept and whom they accept as a credential issuer.

\subsubsection*{Privacy}

In our framework, users can store and manage their identity data independently, without having to rely on a distinguished third party. Communication is designed to be only bilateral between a credential owner and verifier, and only requires occasional, potentially anonymized read queries to a random node on a public blockchain to update schemas, issuers' signing keys, and revocation registries. This architecture prevents third parties from surreptitiously gaining insights into users' comprehensive data, as is the case with federated identity providers (R\,5.1). This is also desirable from a scalability perspective (Expert~G). As a result, users can have different digital identities in different contexts and only need to disclose the data required for a specific situation (Experts~E and F) in accordance with the \emph{need to know} principle.

In the \ac{SSI}-based approach, customers have complete control over the data in their wallets, and customers can decide for themselves whom they wish to share data with (Experts~D,\,E, and F). In this context, the experts also mentioned the possibility of selective disclosure through \ac{ZKP}. The fact that customers no longer have to show all their personal details, but only the relevant data, helps to protect customer privacy through data minimization (R\,5.2) (Experts E and F).

The experts also emphasized that a correlation of data is still possible in the absence of public identifiers on the basis of the available rich data sets of banks and other organizations. However, \ac{SSI}'s goal is not anonymity, as is sometimes suggested, but rather the best possible extent of privacy in each scenario. Since \ac{KYC} procedures seek to build  trust, a large amount of personal information must be revealed. In this context, it is important to note that researchers such as~\citet{lootsma2017blockchain} have emphasized the possibility of harnessing additional potential through \ac{KYC} data by, for instance, connecting them to transactional data. While~\citet{lootsma2017blockchain} have even raised the question of resulting conflicts with customer privacy, such efforts may also lead to an inherent conflict with \ac{SSI} principles. Nonetheless, once personally identifiable information has been received in plain text through a bank, it cannot be hindered in connecting it to other data, also in our approach.

\subsubsection*{User experience}

The \ac{SSI}-based \ac{eKYC} process has the potential to vastly improve the user experience of customers. Much of the current friction, such as entering personal data in an online form, the need to visit a bank, or the need to make a video call with a bank employee for identification process, has been eliminated. Instead, the onboarding process can be carried out on the user's smartphone with just a few steps, for instance by scanning QR~codes and accepting invitation links and proof requests through simple interfaces (R\,6.1).
Because the framework builds on generic and open standards, for which many reference implementations for mobile phones and computer operating systems are available, different user interfaces are realizable (R\,6.2).
Further, customers have a permanent overview of whom they shared data with (Expert~E). A potential problem lies with the customer's full responsibility for data storage and administration (Experts~D,~E, and G) (R\,6.3). Customers must develop an awareness of this so that they realize their responsibility and take appropriate backup and recovery measures to mitigate the consequences of device loss or theft (Experts E and G).

\section{Discussion}
\label{sec:discussion}

As indicated in Section~\ref{sec:evaluation}, our framework can greatly improve \ac{KYC} processes. In particular, efficiency, trust, and privacy seem to benefit from the blockchain-based \ac{SSI} architecture. However, many of the improvements do not specifically relate to the \ac{KYC} case. The trust relationship illustrated in Figure~\ref{fig:SSI_Layers} between a holder of ID attributes, an issuer of documents confirming these attributes, and a verifier is present in multiple domains. Thus, our architecture and its related processes reveal insights into the general design of artifacts in the nascent field of blockchain-based \ac{SSI}, which according to the interviews with the \ac{SSI} experts may translate to many other areas, where the fear of a centralized service provider has so far prevented a more efficient cross-organizational identity management. To elevate our IT artifact for further theoretical discussion, we derived three \acp{DP} that abstract our findings and that seek to guide future research and practice in blockchain-based \ac{SSI}~\citep{gregor2013positioning}. We analyzed codes from the interviews related to our architecture's technical building blocks (e.g., (distributed) ledger, blockchain, \ac{VC}, or storage) to identify commonly proposed design patterns and their justification, and we arrived at three generic principles.

\subsection*{Design principle 1: Utilize blockchain only for public data}
Our research suggests that the absence of a centralized platform operator in the \ac{eKYC} process can enable cooperation between banks. The banks do not have to fear that other banks or even a central \ac{eKYC} utility will receive valuable customer data, which could put them in a disadvantageous position or create new dependencies and lock-in effects. In this context, a distributed ledger is well suited to transparently display public information. On the other hand, owing to their inherent properties -- such as transparency, redundancy, and tamper-resistance -- blockchains are not suitable for storing personal data~\citep{zhang2020privacy}, even in encrypted form~\citep{finck2018blockchains,covidcredentialsinitiative}.
The academic literature often states that credential hashes and peer \acp{DID} also need to be stored on a distributed ledger~\citep{muhle2018survey}, and initial frameworks for \ac{KYC} based on \ac{SSI}~\citep{soltani2018new} have used this approach. However, from a technical perspective and according to the experts, this has no apparent advantages and only carries performance challenges and regulatory risks: Trust in the interaction with a \ac{DID} is established through a \ac{VP}, and \acp{VC}' tamper resistance is established via the issuer's digital signatures, which need to be trusted anyway. This renders on-chain hashes unnecessary~\citep{toth2019self}. Further, it must be assumed that legal persons' \acp{DID} fall under the \ac{GDPR}~\citep{wagner2018self}, and for the aforementioned reasons, they should not be stored on a distributed ledger. Thus, distributed ledgers should only be used in a manner comparable to a \ac{PKI} for \ac{VC} issuers (Experts~E and F) and not for private persons' \acp{DID} and \acp{VC} (Experts~E,\,F, and G). By taking most communication off-chain, as Expert~G mentioned, the proposed architecture and \ac{SSI} could help many blockchain use cases to comply with regulation such as the \ac{GDPR} or \ac{eIDAS} and could resolve privacy issues (Experts~C,\,D,\,E,\,F and G). Regarding performance, the Hyperledger Indy blockchains on which many \ac{SSI} systems rely can handle only a limited number of write transactions~\citep{sedlmeir2020DLPS}; thus, one should design interactions between stakeholders in a blockchain-based \ac{SSI} environment bilaterally if possible, and one should read from a blockchain rather than write to it so as to avoid scalability issues.

To abstract and generalize this observation, we propose that by using \ac{SSI} in processes that require proofs about the possession of certain attributes, organizations should repeatedly request and verify these attributes through bilateral communication channels, instead of storing the required data centrally. As a side effect, this can also help keep data up to date.

\subsection*{Design principle 2: Anticipate an ecosystem of various ledgers}
Our initially designed framework was built on the assumption that financial institutions share a single distributed ledger to create and manage digital identities for \ac{eKYC}. Employing a shared ledger facilitates interoperability on a technical level and concerning governance. However, recent developments in \ac{SSI} practice~\citep{kuperberg2019blockchain} and our interview findings indicate that it is more likely that various distributed ledgers for \ac{SSI} will exist (Experts F and G), similar to the considerable number of today's certificate authorities. Thus, it is important to account for this circumstance and to design blockchain-based \ac{SSI} solutions for various distributed ledgers to achieve interoperability. This can be achieved through adherence to industry standards, which are currently being developed by organizations such as the \ac{W3C}~\citep{w3c2019vc}, as well as by using technical components for interoperability and trust. Universal resolvers -- i.e., identifier resolvers working for a multitude of identifiers such as \acp{DID} on different blockchains and maybe also centralized databases (e.g., provided by certificate authorities) -- may play an important role in this regard and may also increase trust (Experts~D,\,E,\,F, and G). While interoperability is technically achievable without major challenges, the existence of various distributed ledgers may induce governance-level challenges that must also be designed through cross-ledger governance (Expert~G).

\subsection*{Design principle 3: Enable decentralization at the edge}
During the creation of the \ac{SSI}-based \ac{KYC} framework, we encountered some challenges regarding \ac{SSI}-based identity management's user-friendliness. Although managing all identity-relevant data through a single app can boost the straightforward and user-friendly management of identity documents and increase authenticity through hardware-binding or credential-linking, self-managing leads to multiple challenges. For instance, users need support if their devices are lost or stolen. The status quo of central systems must be broken down somewhat here~\citep{wagner2018self}, and users must develop an awareness of their responsibility for their data and must learn to store them accordingly (Experts E and F). On the other hand, one must also find the right balance between decentralized and central solutions~\citep{dunphy2018first}. An example can be the use of cloud storage and cloud agents, which can add value concerning recovery, availability, and security if the agent specializes in this service. On the other hand, these cloud solutions contradict \ac{SSI}'s basic idea of avoiding as many third parties as possible, particularly honey-pots of data, for privacy and security reasons. However, as long as the data are encrypted and the cloud providers cannot access the data, Expert~C sees cloud storage a both a viable and an essential element for enabling good user experience.
To think decentralization to an end and support the autonomy of end users in blockchain-based \ac{SSI} applications, \ac{SSI}-based architectures must ensure that users can store their \acp{VC} on an infrastructure of their choice.

\subsection*{Crossing the chasm: How to bring blockchain-based \ac{SSI} into practice in \ac{KYC} and beyond}

While design artifacts, as the outcome of the DSR process, should have practical impact~\citep{baskerville2018design}, bringing the artifacts into practice requires suitable approaches. One key topic mentioned by multiple experts -- but that does not relate to technical terms and is therefore not a \ac{DP} that we can derive based on our codes related to technical building blocks -- is the adoption of the \ac{SSI}-based \ac{eKYC}. These experts regarded the general adoption of \ac{SSI} technology in the public and private sectors as a major driving force of the practical implementation of our \ac{eKYC} solution. Expert~D called this a \emph{chicken and egg} problem: Since the technology is still very new, there are few credential providers, so the utility for a user is very low; on the other hand, as long as there are only a few users, there is also no major incentive for organizations to act as credential issuers. The more credentials users have, the better such a system can be used (Expert~D).
Network effects can help bridge the gap between early adopters and the widespread use of the technology~\citep{moore1999crossing}. In particular, banks can contribute to this by offering the technology in the \emph{isolated} \ac{KYC} use case, issuing \acp{VC} to contribute to its spread in the mass market. The more that banks accept these credentials, the more attractive the system becomes to customers. In turn, higher usage by customers leads to more incentive for other organizations to accept \acp{VC}. As mentioned, the cooperation of the banks and the creation of shared standards are crucial if this adoption is to become possible.

Because users have full control over their data, \ac{SSI} must address the \ac{GDPR}'s general requirements, such as privacy by design, portability, the right to erasure, transparency, purpose limitation, data minimization, accuracy, storage limitation, and information integrity. Users can get a permanent overview of whom they shared what data with (Expert~E), and these records can help to better implement the right to erasure.
In turn, this may lead to a better adoption of the technology. In fact, the \ac{GDPR}'s strict requirements, which were often criticized for impeding blockchain-related innovation in Europe, may ultimately have turned out to boost innovation, so blockchain's benefits for interoperability can be used for the purposes highlighted in \ac{DP}~1 while avoiding its well-known privacy- and scalability-related challenges.

The interplay between \ac{SSI} and regulation uncovers many other interesting dimensions. For instance, our interviewees suggested that \ac{eIDAS} regulation can facilitate \ac{SSI} while \ac{SSI} can help make the \ac{eIDAS} infrastructure, which so far has been used only moderately, more practicable and valuated~\citep{lyon2019identity}. On the other hand, we saw that \ac{SSI} technically allows for even more privacy than what is required by regulation. However, the \ac{MLA} requires that banks store customer data for five years, creating tension between data protection regulation and the objectives of user control and the prevention of data silos. Thus, \ac{SSI} may even lead to new discussions on where to set the sweet spot between market integrity and privacy.

\section{Conclusion}
\label{sec:conclusion}
In this article, we sought to build a framework to improve on the current shortcomings in the KYC process through an end-to-end digital process that leverages blockchain-based \ac{SSI}. Research on \ac{SSI} is still in its infancy, and little has been published on the design of applications for SSI. \citet{soltani2018new} were the first to explore this topic in the context of \ac{KYC}, covering the onboarding process and technically evaluating their solution. Building on this valuable work, we extended the scope and emphasized banks' requirements. We used a \ac{DSR} approach based on~\citet{peffers2007design}, designing and evaluating a framework for \ac{KYC} processes built on blockchain-based \ac{SSI}, including a generic architecture and process design. Since we face a low solution maturity in the innovative field of blockchain and \ac{SSI}, and high application domain maturity in the domain of \ac{KYC}, we provided an \textit{improvement} in the context of \ac{DSR}~\citep{gregor2013positioning}. Our evaluation suggests that our design can significantly contribute to a more efficient \ac{KYC} process that also addresses the other requirements of stakeholders. Thus, we are confident that we have accomplished our research objective.

Besides the conceptualized and evaluated architecture and set of processes~\citep{gregor2013positioning} for the \ac{KYC} case, we made three primary contributions to the academic body of knowledge. First, our examination revealed the challenges of using \ac{DLT} for the exchange of personal data generally and particularly for digital identity management systems. We also showed how these problems can be solved by using \ac{SSI} on top of the blockchain layer, thereby leveraging the advantages typically associated with blockchain technology while avoiding its well-known issues with scalability and privacy. Second, we revealed the implications of designing \ac{SSI}-based solutions built on blockchain in the context of \ac{KYC} by deriving three \acp{DP}, which allowed us to elevate our IT artifact for more abstract and generalizable theoretical discussion~\citep{gregor2013positioning}. Third, we offered suggestions for relevant future research on blockchain and \ac{SSI}, enabling researchers to base their work on our results and thus generate additional knowledge~\citep{vom2020special}.

\ac{DSR} should also inform practice to advance a specific domain through IT~\citep{gregor2013positioning}. Our conception and evaluation of the \ac{SSI}-based \ac{KYC} framework will provide practitioners with valuable insights regarding design choices, \ac{DLT}'s role, the intricacies of regulation, and related challenges and opportunities for banks and customers. Our results indicate that \ac{SSI}-based \ac{eKYC} processes can reduce cost and time expenditures and contribute to better user experiences and increased security during the \ac{KYC} process. We demonstrated how the use of \ac{SSI} can positively impact the different onboarding processes and their interplays with an existing \ac{SSI} ecosystem. However, we illustrated that there are further conceptual challenges to be solved before SSI is used in real systems and settings, especially regarding the necessary governance frameworks and a more detailed regulatory analysis.
While our research suggests synergies between \ac{SSI} and regulation, challenges remain, especially to establish a general \ac{SSI}-based ecosystem and to make \ac{SSI} as user-friendly as possible without sacrificing privacy and security.

Like most research, our study has limitations. Our framework has not yet been used in practice and therefore lacks an evaluation in a real-world setting. However, by applying a rigorous research design and obtaining practitioner feedback, we sought to address this shortcoming. Further, although we described the necessary and central elements of an additionally required governance framework, its concrete implementation remains open. In particular, details regarding the cooperation of banks in the \ac{KYC} process, the creation of shared standards, and the responsible parties for operating the blockchain in the case of using a permissioned network must be clarified. This opens various promising avenues for future research. In particular, the \ac{KYC} process is often relevant not only in the banking environment but also in other domains such as insurance, and further objectives may be necessary to address this aspect. Further, although our interviews with experts  confirmed~\citeauthor{moyano2017kyc}'s~[\citeyear{moyano2017kyc}] findings that there is sufficient trust between banks that collaboration on \ac{eKYC} is possible despite competition on a suitable IT infrastructure, this only represents the perspective of practitioners and researchers in central and northern European countries. Nonetheless, there is also a promising development that may make \ac{SSI}-based \ac{KYC} and our findings considerably more far-reaching and applicable in contexts in which this trust is missing: In emerging \ac{SSI} ecosystems in North America and Europe, governments are starting to explore the impacts of issuing certificates such as driver's licenses and ID cards in the form of \acp{VC} that can be leveraged by the public and private sector. Besides increasing the efficiency of digital identification and authentication, a digital ID is expected to contribute to substantial reductions in financial crime~\citep{ft2021fraud}. In this context, an experimental clause was recently adopted in Germany's parliament that explicitly allows banks to perform \ac{KYC} based on an ID card in the form of a \ac{VC}~\citep{bankenverband2021ssi,bundestag2021experimental}.

We are confident that we have derived guidelines and \acp{DP} that generalize to these promising developments and to other sectors, and that also provide guidance on how to make blockchain-based \ac{SSI} compatible with the needs of businesses and regulatory restrictions. More efficient, reusable processes, specifically relating to identity management, are needed in both business and the public sector, but the risks associated with central providers have so far prevented general services from providing these capabilities. Blockchain-based \ac{SSI} can address the need for a general service for data that is non-competitive, since the same data are needed and used by all, and yet comply with customer data privacy expectations and regulations. On the other hand, our \ac{DP} of minimal involvement of the blockchain, specifically not storing natural persons' \acp{DID} or even \acp{VC} on a blockchain, translates to applications of \ac{SSI} generally, and we are eager to see more use cases being built on this technology stack. Given current efforts by the Verifiable Organizations Network in Canada, the ambitious goal of having 10 different pilots that leverage blockchain-based \ac{SSI} by the end of 2021 in Germany, and several \ac{SSI} smartphone wallets that are already available, we are confident that this time, the promises of blockchain-technology to revolutionize digital ID management can be fulfilled, although blockchain's role will be much more restricted than what early research suggested.

\end{document}